\documentclass[11pt,a4paper]{article}
\usepackage{jheppub}\pdfoutput=1
\usepackage[T1]{fontenc}
\usepackage{graphicx}
\usepackage{amsmath}
\usepackage{amsmath}
\usepackage{xspace}
\usepackage{subfig}
\usepackage{color}
\usepackage[utf8]{inputenc}
\usepackage{commath}
\usepackage{slashed}
\usepackage{fancyvrb}
\usepackage{multirow}
\usepackage[utf8]{inputenc}
\usepackage[english]{babel}
\usepackage{amsfonts}
\usepackage{amssymb}
\usepackage{cancel}
\newcommand{\lsim}{\mbox{\raisebox{-.6ex}{~$\stackrel{<}{\sim}$~}}}
\newcommand{\gsim}{\mbox{\raisebox{-.6ex}{~$\stackrel{>}{\sim}$~}}}

\title{Singlet-doublet fermion Dark Natter with Dirac neutrino mass, $(g-2)_\mu$ and $\Delta N_{\rm eff}$}

\author[a]{Debasish Borah}
\emailAdd{dborah@iitg.ac.in}
\affiliation[a]{Department of Physics, Indian Institute of Technology Guwahati, Assam 781039, India}

\author[b]{Satyabrata Mahapatra}
\emailAdd{satyabrata@g.skku.edu}
\affiliation[b]{Department of Physics and Institute of Basic Science, Sungkyunkwan University, Suwon 16419, Korea}

\author[c]{Dibyendu Nanda}
\emailAdd{dnanda@kias.re.kr}
\affiliation[c]{School of Physics, Korea Institute for Advanced Study, Seoul 02455, Korea}

\author[d]{Sujit Kumar Sahoo}
\emailAdd{ph21resch11008@iith.ac.in}
\affiliation[d]{Department of Physics, Indian Institute of Technology Hyderabad, Kandi, Sangareddy 502285, Telangana, India}

\author[d]{Narendra Sahu}
\emailAdd{nsahu@phy.iith.ac.in}

\abstract
{We study the possibility of generating light Dirac neutrino mass via scotogenic mechanism where singlet-doublet fermion Dark Matter (DM) plays non-trivial role in generating one-loop neutrino mass, anomalous magnetic moment of muon: $(g-2)_\mu$ as well as additional relativistic degrees of freedom $\Delta{N_{\rm eff}}$ within reach of cosmic microwave background (CMB) experiments. We show that the Dirac nature of neutrinos can bring interesting correlations within the parameter space satisfying the $\left(g-2\right)_\mu$ , DM relic density and the effective relativistic degrees of freedom $\Delta{N_{\rm eff}}$. While we stick to thermal singlet-doublet DM with promising detection prospects, both thermal and non-thermal origin of $\Delta{N_{\rm eff}}$ have been explored. In addition to detection prospects of the model at DM, $(g-2)_\mu$ and other particle physics experiments, it remains verifiable at future CMB experiments like CMB-S4 and SPT-3G.}

\begin{document}
	\newenvironment{myenv}[1]
	{\everydisplay{\setlength\displaywidth{#1}}\begin{eqnarray}}
	{\end{eqnarray}\ignorespacesafterend}
	
	\maketitle
	
	\section{Introduction}
	\label{intro}

The mysteries of the Universe have always motivated us to explore the fundamental aspects of physics beyond the standard model(SM) of particle physics. The prime among them are understanding the dark matter (DM) component, the origin of neutrino mass and the anomalous magnetic moment of muon: $(g-2)_\mu$, to count a few, that challenge the phenomenological success of the SM.
DM is particularly puzzling because we can only detect its presence through its gravitational effects on visible matter and the structure of the Universe. Studies using satellites like WMAP and PLANCK~\cite{Planck:2018vyg, WMAP:2012nax} measuring the anisotropies in the cosmic microwave background(CMB) have shown that DM makes up about 85\% of all matter and approximately 27\% of the Universe's total energy density. The current abundance of DM is frequently expressed in terms of the density parameter $\Omega_{\rm DM}$ and the normalized Hubble parameter $h$, defined as the Hubble parameter divided by $100$ km ${\rm s}^{-1} {\rm Mpc}^{-1}$, yielding $\Omega_{\rm DM}h^{2} = 0.120\pm0.001$ at a $68\%$ C.L.. Despite its significant role in cosmology, DM remains elusive, as no known SM particles fit the criteria of being DM. This has led researchers to consider new theories beyond the SM, with the weakly interacting massive particle (WIMP) model being one of the most studied. The WIMP model suggests that DM particles could have masses and interaction strengths similar to those of electroweak particles, potentially explaining the observed abundance of DM through a process known as thermal freeze-out. However, despite extensive searches, no direct evidence of DM has been found, placing strict limits on the possible interactions between DM and ordinary matter. A comprehensive assessment of WIMP-type DM models is available in a recent review~\cite{Lee:1977ua, Griest:1989wd}.

In addition to DM, SM also fails to explain the origin of neutrino mass and mixing, which has been confirmed by neutrino oscillation experiments~\cite{Super-Kamiokande:1998kpq, SNO:2001kpb, DoubleChooz:2011ymz, DayaBay:2012fng, RENO:2012mkc}. Neutrino oscillation data only captures the differences in the squared masses of neutrinos( $\Delta m^2_{\rm atm}=(2.358-2.544)\times10^{-3}~{\rm eV}^2$ and $\Delta m^2_{\rm sol.}=(6.79-8.01)\times10^{-5}~{\rm eV}^2$), while cosmological data place limits on their absolute mass scale ($|\sum_i m_{\nu_i}|\leq0.12$ eV)~\cite{ParticleDataGroup:2022pth}. However the question of whether neutrinos are Dirac or Majorana particles remains open. While neutrino oscillation experiments cannot settle this, searches for neutrino-less double beta decay might confirm the Majorana nature of neutrinos. But so far, no evidence has been found leaving the Majorana nature of light neutrinos unverified. This has kindled increased interest in exploring the plausibility of light Dirac neutrinos. While traditional Dirac neutrino mass models built on the seesaw mechanism have been discussed in~\cite{Biswas:2021kio, Borah:2022obi, Biswas:2022vkq}, alternative scenarios outlining the possibility of light Dirac neutrino mass can be found in~\cite{Ma:2016mwh,Yao:2017vtm,Carvajal:2018ohk,Jana:2019mgj,Nanda:2019nqy, Borah:2020jzi,  Biswas:2022fga,Wang:2016lve} and related references.

Another unresolved issue within the SM is the anomalous magnetic moment of the muon defined as $a_\mu=(g-2)_\mu/2$. The quantity $\Delta a_\mu= a^{\rm exp}_\mu - a^{\rm SM}_\mu$, measures the difference between the muon's measured magnetic moment and the value predicted by the SM. Recent experiments have reported a discrepancy between the measured and predicted values of $(g-2)_\mu$, suggesting the possibility of new physics beyond the SM~\cite{ Aoyama:2012wk,Aoyama:2019ryr,Czarnecki:2002nt,Gnendiger:2013pva,Davier:2017zfy,Keshavarzi:2018mgv,Colangelo:2018mtw,Hoferichter:2019mqg,Davier:2019can,Keshavarzi:2019abf,Kurz:2014wya,Melnikov:2003xd,Masjuan:2017tvw,Colangelo:2017fiz,Hoferichter:2018kwz,Gerardin:2019vio,Bijnens:2019ghy,Colangelo:2019uex,Blum:2019ugy,Colangelo:2014qya,Cherchiglia:2023utd,Crivellin:2021rbq, Crivellin:2018qmi}. The 2021 analysis conducted by the $(g-2)_\mu$ collaboration, with the Fermilab's E989 experiment result in conjunction with the prior results from Brookhaven, revealed a discrepancy of $4.2\sigma$. A more recent analysis by the same collaboration \cite{Muong-2:2023cdq} has yielded $ \Delta a_\mu = 249 (48)\times 10^{-11}$, indicating a discrepancy of $5.1\sigma$ from the SM predicted value.  However, it is important to note that, due to the non-perturbative nature of the low-energy strong interaction, the uncertainty in $a^{\rm SM}_\mu$ is primarily dominated by contributions from hadronic vacuum polarization (HVP), which is difficult to calculate precisely. Despite recent measurements~\cite{Aoyama:2020ynm,CMD-3:2023alj} that have reduced this uncertainty, there is still no conclusive evidence, leaving room for alternative BSM explanations. A review of such new physics explanations for $(g-2)_\mu$ can be found in \cite{Jegerlehner:2009ry, Lindner:2016bgg, Athron:2021iuf}.

Although the origin of the aforementioned puzzles remain unknown, in this paper we aim to propose a concise model that explains them within a common framework. We explore a WIMP-like fermionic DM scenario, involving a vector-like singlet and a doublet~\cite{Bhattacharya:2015qpa,Bhattacharya:2018fus}. The rationale for exploring this particular fermionic configuration is well-established as in
singlet-doublet setups, there is mixing between the neutral component of the doublet and the singlet through the Yukawa interaction and DM manifests as a mixed state~\cite{Banerjee:2016hsk, DuttaBanik:2018emv, Horiuchi:2016tqw, Restrepo:2015ura, Badziak:2017the, Betancur:2020fdl, Abe:2017glm, Abe:2019wku, Barman:2019tuo, Calibbi:2018fqf, Fraser:2020dpy, Freitas:2015hsa, Cynolter:2015sua, Calibbi:2015nha, Abe:2014gua, Cheung:2013dua, Cohen:2011ec, Dutta:2021uxd, Ghosh:2023dhj, Borah:2021khc, Bhattacharya:2021ltd, Bhattacharya:2018cgx, Bhattacharya:2015qpa, Bhattacharya:2018fus, Bhattacharya:2016rqj, Bhattacharya:2017sml, Dutta:2020xwn, Konar:2020wvl, Konar:2020vuu, Borah:2022zim}, satisfying appropriate relic density across a broad range of masses while adhering to direct search limits, which is not the case for purely a singlet or a doublet scenario. Additionally, we extend this model with one scalar and three right-handed neutrinos ($\nu_{R_i}$) to incorporate the Dirac mass of light neutrinos in a scotogenic setup, offering complementary cosmological probes in CMB measurements in contrast to the singlet-doublet fermionic DM setups considered in literature in connection to Majorana neutrino masses~\cite{Dutta:2020xwn, Borah:2021rbx, Borah:2022zim, Konar:2020wvl, Konar:2020vuu, Bhattacharya:2017sml,Bhattacharya:2016rqj}.

The presence of right handed neutrinos in our model potentially introduce extra light degrees of freedom. Current CMB experiments have placed stringent constraints on effective light degrees of freedom during the recombination era ($z\sim 1100$), yielding ${\rm N_{eff}= 2.99^{+0.34}_{-0.33}}$ at $2\sigma$ aligning with SM predictions of ${\rm N}^{\rm SM}_{\rm eff}=3.046$ ~\cite{Planck:2018vyg, WMAP:2012nax}. Future experiments like CMB-S4 aim for unprecedented sensitivity, targeting $\Delta {\rm N}_{\rm eff} = 0.06$ at 2$\sigma$ CL~\cite{Abazajian:2019eic}. This precision measurement can scrutinize scenarios featuring light degrees of freedom that we have. 
Due to the minimality of the framework, which offers a unified resolution to the aforementioned challenges, it reveals a strong correlation among the model parameters, making the model extremely predictive at various present and future experiments.

The rest of the paper is organized as follows.	
 In section \ref{section:model}, we discuss a most minimal extended version of the singlet-doublet fermionic model to incorporate radiative Dirac neutrino mass. In section \ref{section:numass}, we show the generation of Dirac neutrino mass at one-loop level. In section \ref{section:mug2}, we analyse the bounds from $(g-2)_\mu$ on our model parameters followed by the lepton flavour violation constraint. In section \ref{section:DMpheno}, we discuss the relic density of DM and direct detection constraints. In section \ref{section:delneff}, we discuss the details of $\Delta N_{\rm eff}$ and finally conclude in section \ref{section:conclusion}.

 
\section{The Model}
\label{section:model}
As discussed in the introduction, to generate the Dirac mass of light neutrinos in a scotogenic setup we augment 
the SM fermion content by adding 2 generations of vector-like fermion doublet $\Psi=(\psi^{0} ~\psi^{-})^{T}$ (with hypercharge $Y = -1$, where we use the electromagnetic charge definition $Q = T_3 + Y /2$) and 2 generations of vector-like singlet fermion $\chi$ with hypercharge $Y=0$, together with three copies of right handed neutrinos $\nu_R$. 
In addition to the singlet-doublet fermionic extension of SM, we introduce another singlet complex scalar $\phi(=\frac{\phi_1+i\phi_2}{\sqrt{2}})$. With this particle content, we introduce a discrete $Z_4$ symmetry, where this new symmetry is responsible for forbidding the tree level realization of Dirac neutrino mass and the Majorana mass of $\nu_R$. The charge assignments of the particles under the imposed symmetries are shown in Table.~\ref{tab1}.

	\begin{table}[h!]
		\small
		\begin{center}
			\begin{tabular}{||@{\hspace{0cm}}c@{\hspace{0cm}}|@{\hspace{0cm}}c@{\hspace{0cm}}|@{\hspace{0cm}}c@{\hspace{0cm}}|@{\hspace{0cm}}c@{\hspace{0cm}}||}
				\hline
				\hline
				\begin{tabular}{c}
					{\bf ~~~~ Gauge~~~~}\\
					{\bf ~~~~Group~~~~}\\ 
					\hline
					
					$SU(2)_{L}$\\ 
					\hline
					$U(1)_{Y}$\\ 
					\hline
					$Z_4$\\
				\end{tabular}
				&
				&
				\begin{tabular}{c|c|c|c|c}
					\multicolumn{5}{c}{\bf Fermion Fields}\\
					\hline
					~ $L_L$&$l_R$&$\Psi$ & $\chi$~~~& $\nu_R$ \\
					\hline
					
					$2$&$1$&$2$ & $1$&$1$\\
					\hline
					$-1$&$-2$&$-1$ & $0$&$0$\\
					\hline
					$-i$&$-i$&$-1$& $-1$&$i$ \\
				\end{tabular}
				&
				\begin{tabular}{c|c}
					\multicolumn{2}{c}{\bf Scalar }\\
					\hline
					~$H$&$\phi$\\
					\hline
					$2$&$1$ \\
					\hline
					$1$&$0$ \\
					\hline
					$+1$&$i$ \\
				\end{tabular}\\
				\hline
				\hline
			\end{tabular}
			\caption{Particles and their
				charge assignments in our setup.}
			\label{tab1}
		\end{center}    
	\end{table}
	
The relevant Lagrangian of the model guided by the imposed symmetry is given by:
\begin{eqnarray}
\mathcal{L} &\supset& i \overline{\Psi} \gamma^\mu D_\mu \Psi + i \overline{\chi} \gamma^\mu \partial_\mu \chi - M_{\Psi} \overline{\Psi} \Psi -M_{\chi} \overline{\chi} \chi 
\nonumber\\&{}&-y\overline{\Psi} \Tilde{H} \chi - \lambda_{\psi}\overline{L} \phi \Psi-\lambda_{\chi} \overline{\nu_{R}} \phi^\dagger  \chi + h.c.,
\label{eq:lag}
\end{eqnarray}
where $D_\mu=\partial_\mu-g_1\frac{\tau_i}{2} W^i_\mu-g_2\frac{Y}{2} B_\mu$. For simplicity, we suppress the generation indices in the Lagrangian.
The scalar potential  involving $H$ and $\phi$ which satisfy $Z_4$ symmetry is given by:
\begin{eqnarray}
V_{H,\phi} &=&  -\mu^2_H (H^\dagger H) + \lambda_H (H^\dagger H)^{2} \nonumber\\& +& {M^2_\phi} (\phi^\dagger \phi) + {\lambda_{\phi}} (\phi^\dagger \phi)^2  +\lambda_{\phi H} (\phi^\dagger \phi) (H^{\dagger} H)
\label{eq:scalarL}
\end{eqnarray}
In our setup, we choose $M^2_\phi > 0$ such that $\phi$ doesn't acquire any vacuum expectation value (VEV), thus forbidding any mixing between $\nu_{R}$ and $\chi$. Moreover, the SM Higgs $H$ is inert under $Z_4$ symmetry and hence can not lead to spontaneous $Z_4$ breaking together with electroweak symmetry breaking.

In order to generate the Dirac mass of neutrinos, one has to break $Z_4$ symmetry. In order to do so, we add a soft term $\frac{1}{2}\mu^2_{\phi}\left(\phi^2 + (\phi^\dagger)^2\right)$ in the scalar potential which breaks $Z_4$ symmetry explicitly.
After electroweak symmetry breaking, SM Higgs acquires a VEV. In presence of the soft term, the masses of $\phi_1$ and $\phi_2$ can be written as,
\begin{eqnarray}
    M^2_{\phi_1}&=& M^2_\phi+\mu^2_\phi +\frac{1}{2}\lambda_{\phi H} v^2 \nonumber \\
    M^2_{\phi_2}&=& M^2_\phi-\mu^2_\phi +\frac{1}{2}\lambda_{\phi H} v^2
\end{eqnarray}
and the mass squared difference, $\Delta M^2_{\phi} = M^2_{\phi_1}-M^2_{\phi_2} = 2 \mu^2_{\phi}$. We will show that $\Delta M^2_{\phi}$ is responsible for generating Dirac mass of neutrino at one-loop level in Sec. \ref{section:numass}.
The Higgs VEV also leads to mixing between the neutral component of doublet $\psi^0$ and $\chi$, giving rise to two singlet-doublet admixed mass eigenstates. The mass terms for these fields can then be written together as follows:
\begin{eqnarray}
-\mathcal{L}^{VF}_{mass}&=&M_\Psi\overline{\psi^0}\psi^0+M_\Psi{\psi^+}\psi^-+M_{\chi} \overline{\chi}\chi \nonumber\\&+&  \frac{y v}{\sqrt2}~ \overline{\psi^0}~\chi 
+\frac{y v}{\sqrt2} ~\overline{\chi}~\psi^0 \nonumber \\
&=&
\overline{\left(\begin{matrix}
		\psi^0 & \chi 
		\end{matrix}\right)}
{\left(\begin{matrix}
		M_{\psi} & y v/\sqrt2\\
		y v/\sqrt2 & M_\chi
		\end{matrix}\right)}
{\left(\begin{matrix}
		\psi^0 \\ \chi
		\end{matrix}\right)}\nonumber\\
&+&M_\Psi {\psi^+}\psi^- . \nonumber\\
\label{vlfyuk}
\end{eqnarray}
	
Denoting the mass eigenstates as $\chi_1$ and $\chi_2$ and $\theta$ to be the mixing angle, the flavour basis, $\left(
	\psi^0~~~ \chi 
	\right)^T$ is related to the physical basis, $\left(
	\chi_1 ~~~  \chi_2 
	\right)^T$ through the  orthogonal transformation:
	\begin{eqnarray}
	\left(\begin{matrix}
	\psi^0 \\ \chi
	\end{matrix}\right) 
	= \mathcal{O} \left(\begin{matrix}
	\chi_1 \\  \chi_2 
	\end{matrix}\right)=\left(\begin{matrix}
	\cos\theta & -\sin\theta \\
	\sin\theta & \cos\theta
	\end{matrix}\right) 
	\left(\begin{matrix}
	\chi_1 \\  \chi_2 
	\end{matrix}\right),
	\end{eqnarray}
	where the mixing angle is given by:
	\begin{eqnarray}\label{eq:mixang}
	\tan{2\theta}=  \frac{\sqrt2 y v}{M_\Psi-M_{\chi}} .
	\end{eqnarray}

	
From Eq.~(\ref{vlfyuk}),  we get the mass eigenvalues of the physical states to be:
\begin{eqnarray}
M_{\chi_1}&=&M_{\Psi} \cos ^2 \theta + \frac{yv}{\sqrt{2}}\sin 2\theta + M_\chi \sin ^2 \theta,~~ \nonumber\\
M_{\chi_2}&=&M_{\Psi} \sin ^2 \theta - \frac{yv}{\sqrt{2}}\sin 2\theta + M_\chi \cos ^2 \theta
\end{eqnarray}
with a mass-splitting $\Delta M=M_{\chi_1}-M_{\chi_2}$.\\

One important thing to note is that the Eqs. \eqref{eq:lag} and \eqref{eq:scalarL} possess an unbroken $Z_2$ symmetry under which $\Psi, \chi,\phi$ are odd while all other particles are even. This remnant symmetry survives even after $Z_4$ is softly broken by the quadratic terms mentioned above. While our choice of soft-breaking terms has resulted in this remnant symmetry, one can also impose it as an additional symmetry while keeping the phenomenology unaltered. We assume a mass hierarchy: $ M_{\chi_1} > M_{\phi_1} > M_{\phi_2} > M_{\chi_2}$. As a result, the lightest $Z_2$ odd particle $\chi_2$, which is dominantly a singlet fermion $\chi$ with a small admixture of the doublet fermion $\Psi$, behaves as a DM.
In Eq. \eqref{eq:lag}, the term $\overline{L}\phi\Psi$ is responsible for $(g-2)_\mu$, the details of which are discussed in Sec. \ref{section:mug2}.\\

In order to satisfy the neutrino oscillation data, a minimum of two generations of singlet-doublet fermions are required. The consideration of multiple generations of singlet-doublet fermions is also motivated in the context of addressing the CDF-II W-mass anomaly, as explored recently in~\cite{Borah:2022zim}. However, in our present work, the role of heavier generation of singlet-doublet fermions is only to generate the correct neutrino data. For rest of the phenomenology discussed here, we focus only on the lighter generation of singlet-doublet fermions. This choice is justified by considering a mass difference of the order of $\mathcal{O}(10^3)$ GeV between the two generations. Consequently, the presence of the heavier generation has negligible impact on the DM phenomenology and the $\Delta N_{\rm eff}$ analysis, as we disregard inter-generational mixing among these dark fermions.
Furthermore, assuming maximal mixing between the singlet and doublet fermions of the heavier generation allows us to fine-tune the corresponding couplings to leptons to smaller values, while being consistent with the neutrino data. This ensures that any additional contribution from the heavier generation of singlet-doublet fermions to $(g-2)_\mu$ and lepton flavor violation (LFV) processes remain suppressed when compared to the contributions from the lighter generation.\\

Thus the relevant parameters of the model are the masses $\{M_{\chi_2},\Delta M(=M_{\chi_1}-M_{\chi_2}), M_{\psi^{-}},\\M_{\phi_1},\Delta M_1(=M_{\phi_2}-M_{\chi_2})\}$, the mixing angle $(\sin\theta)$, and the couplings $\{\lambda_\psi$, $\lambda_\chi\}$. These parameters collectively influence various aspects including the neutrino mass generation, the contribution to muon magnetic moment, DM annihilation and co-annihilation dynamics as well as the thermalisation of $\nu_{R}$ acting as dark radiation and providing additional complementary cosmological probe for this scenario in terms of $\Delta N_{\rm eff}$, which we discuss in details in the subsequent sections of this article.
	
\section{Dirac Neutrino Mass}
\label{section:numass}

\begin{figure}[h]
		\centering
		\includegraphics[scale=0.1]{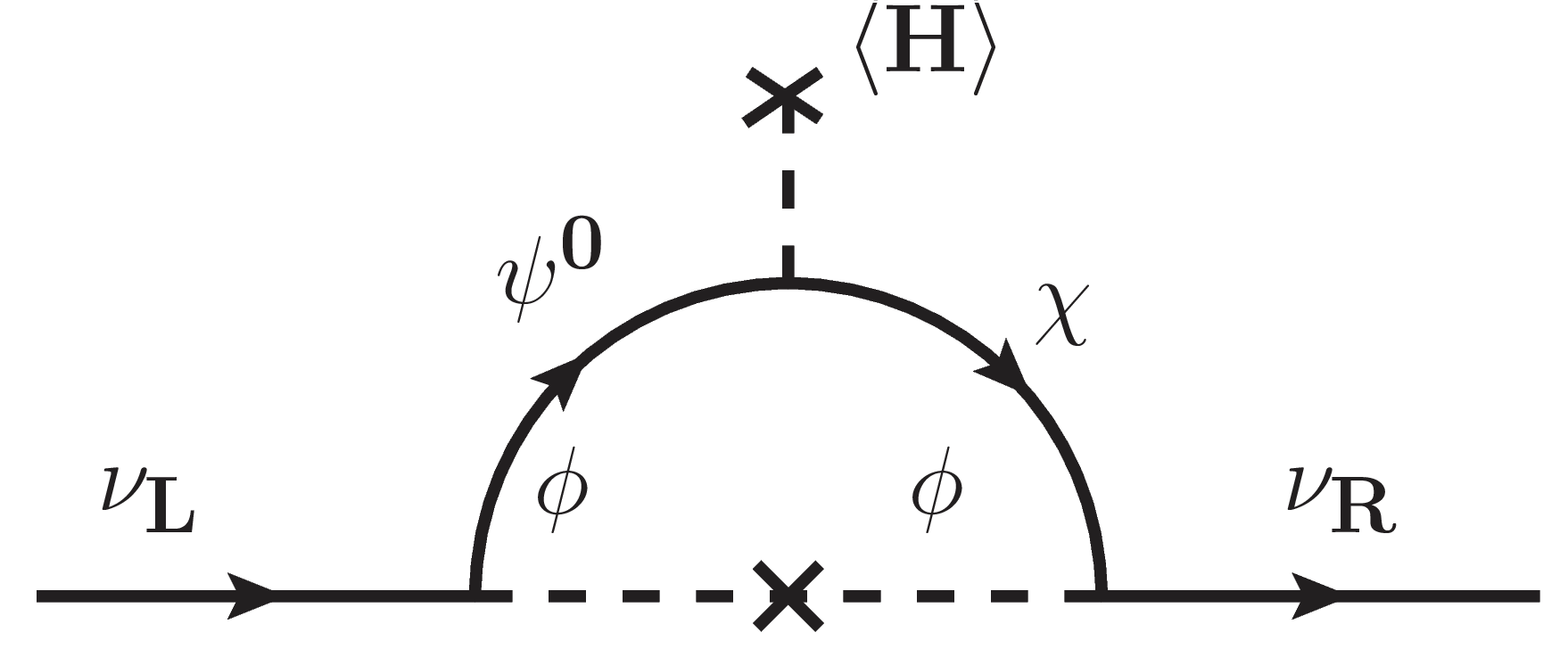}
		\caption{Dirac Neutrino Mass generation at one-loop level.}
		\label{fig:Dirac_neutrino_mass}
\end{figure}

As discussed in the previous section, the $Z_4$ symmetry is explicitly broken by the term $\frac{1}{2}\mu^2_{\phi}\left(\phi^2 + (\phi^\dagger)^2\right)$. This leads to the realisation of Dirac neutrino mass operator $\overline{L}\Tilde{H}\nu_{R}$ at one-loop level with the singlet-doublet fermions($\psi^0,{\chi}$) and the singlet scalar $\phi$ running in the loop as shown in Fig.~\ref{fig:Dirac_neutrino_mass}. 

The one-loop light neutrino mass can be estimated to be :
\begin{eqnarray}\label{eq:nu_mass}
     (M_{\nu})_{\alpha\beta}&=& \frac{\mu^2_{\phi}}{16\pi^2}\sum_i \left({(\lambda_{\psi})}_{i\alpha}\right)^T\left(\Delta M \sin (2\theta) ~F(M_{\chi_1},M_{\chi_2},M_{\phi_1},M_{\phi_2})\right)_i I_{ii}{(\lambda_{\chi})}_{i\beta}
 \end{eqnarray}
where $i$ represents the generation index of singlet-doublet fermion, $\alpha$ and $\beta$ represent lepton flavour indices, $I$ is the $2\times2$ identity matrix and $F(M_{\chi_1},M_{\chi_2},M_{\phi_1},M_{\phi_2})$ is the loop factor and is given by,
{\scriptsize{
 \begin{eqnarray}
     F(M_{\chi_1},M_{\chi_2},M_{\phi_1},M_{\phi_2})&=&\frac{\left(M_{\chi _1}-4 M_{\chi _2}\right) M_{\chi _2}^3 \log \left(\frac{M_{\chi _2}^2}{M_{\chi _1}^2}\right)}{\left(M_{\chi _1}^2-M_{\chi _2}^2\right) \left(M_{\phi _1}^2-M_{\chi _2}^2\right) \left(M_{\phi _2}^2-M_{\chi _2}^2\right)}+  \\
     &{}& \frac{1}{M_{\phi _1}^2-M_{\phi _2}^2}\left[\frac{M_{\phi _1}^2 \left(M_{\chi _1} M_{\chi _2}-4 M_{\phi _1}^2\right) \log \left(\frac{M_{\phi _1}^2}{M_{\chi _1}^2}\right)}{\left(M_{\chi _1}^2-M_{\phi _1}^2\right) \left(M_{\phi _1}^2-M_{\chi _2}^2\right)}-\frac{M_{\phi _2}^2 \left(M_{\chi _1} M_{\chi _2}-4 M_{\phi _2}^2\right) \log \left(\frac{M_{\phi _2}^2}{M_{\chi _1}^2}\right)}{\left(M_{\chi _1}^2-M_{\phi _2}^2\right) \left(M_{\phi _2}^2-M_{\chi _2}^2\right)}\right] \nonumber
 \end{eqnarray}
 }}

For two generations of singlet-doublet fermion, the Eq. \eqref{eq:nu_mass} can be written in a matrix form as,
\begin{eqnarray}\label{eq:nu_matrix}
\left(M_\nu\right)_{3\times3}=
    \left(\begin{matrix}
	M_{ee} & M_{e\mu} & M_{e\tau} \\
	M_{\mu e} & M_{\mu\mu} & M_{\mu\tau} \\
    M_{\tau e} & M_{\tau\mu} & M_{\tau\tau}
	\end{matrix}\right) &=& \sum_{i=1}^2 f^{i}\left(\begin{matrix}
	{\lambda_\psi}_{i e} {\lambda_\chi}_{i e} & {\lambda_\psi}_{i e} {\lambda_\chi}_{i \mu} & {\lambda_\psi}_{i e} {\lambda_\chi}_{i \tau} \\
	{\lambda_\psi}_{i \mu} {\lambda_\chi}_{i e} & {\lambda_\psi}_{i \mu} {\lambda_\chi}_{i \mu} & {\lambda_\psi}_{i \mu} {\lambda_\chi}_{i \tau} \\
    {\lambda_\psi}_{i \tau} {\lambda_\chi}_{i e} & {\lambda_\psi}_{i \tau} {\lambda_\chi}_{i \mu} & {\lambda_\psi}_{i \tau} {\lambda_\chi}_{i \tau}
	\end{matrix}\right) \nonumber \\
 &=& \sum_{i=1}^2 \left(\begin{matrix}
     {\lambda_\psi}_{i e} \\
     {\lambda_\psi}_{i \mu} \\
     {\lambda_\psi}_{i \tau}
 \end{matrix}\right) f^i \left(\begin{matrix}
     {\lambda_\chi}_{i e} & {\lambda_\chi}_{i \mu} & {\lambda_\chi}_{i \tau}
 \end{matrix}\right)
\end{eqnarray}
where $f^{i} = \frac{\mu^2_{\phi}}{16\pi^2}{\left(\Delta M^i \sin2\theta^i ~F(M^i_{\chi_1},M^i_{\chi_2},M_{\phi_1},M_{\phi_2})\right)}$.\\
{\scriptsize{
\begin{table}[h!]
\begin{center}
\begin{tabular}{||@{\hspace{0cm}}c@{\hspace{0cm}}||@{\hspace{0cm}}c@{\hspace{0cm}}||}
\hline
\hline
{\bf Set 1} & {\bf Set 2}\\
\hline
\hline
    \begin{tabular}{ c||c|c } 
    Parameters & $1^{\rm st}$ Generation & $2^{\rm nd}$ Generation \\
    \hline
    \hline
    $\sin\theta$ & $5.8\times 10^{-8}$ & $0.7$\\
    \hline
    $M_{\chi_1}$ (GeV)& $519$ & $3638$ \\
    \hline
    $M_{\chi_2}$ (GeV) & $180$ & $2638$ \\
    \hline
    ${\lambda_\psi}_e$ & $1.38\times 10^{-7}$ & $1.64\times 10^{-7}$ \\
    \hline
    ${\lambda_\psi}_\mu$ & $0.0358$ & $0.0031$ \\ 
    \hline
    ${\lambda_\psi}_\tau$ & $0.0075$ & $0.0090$ \\ 
    \hline
    ${\lambda_\chi}_e$ & $0.0264$ & $4.96\times 10^{-9}$ \\ 
    \hline
    ${\lambda_\chi}_\mu$ & $0.0051$ & $7.40\times 10^{-8}$ \\ 
    \hline
    ${\lambda_\chi}_\tau$ & $2.84$ & $5.05\times 10^{-9}$ \\ 
    \end{tabular}
&
    \begin{tabular}{ c||c|c } 
    Parameters & $1^{\rm st}$ Generation & $2^{\rm nd}$ Generation \\
    \hline
    \hline
    $\sin\theta$ & $0.0053$ & $0.7$\\
    \hline
    $M_{\chi_1}$ (GeV)& $867$ & $3207$ \\
    \hline
    $M_{\chi_2}$ (GeV) & $859$ & $2207$ \\
    \hline
     ${\lambda_\psi}_e$ & $8.30\times 10^{-7}$ & $5.97\times 10^{-5}$ \\
    \hline
    ${\lambda_\psi}_\mu$ & $0.0485$ & $1.64\times 10^{-5}$ \\ 
    \hline
    ${\lambda_\psi}_\tau$ & $0.0295$ & $0.04$ \\ 
    \hline
    ${\lambda_\chi}_e$ & $0.0006$ & $4.87\times 10^{-8}$  \\ 
    \hline
    ${\lambda_\chi}_\mu$ & $4.98\times 10^{-4}$ & $1.00\times 10^{-9}$ \\ 
    \hline
    ${\lambda_\chi}_\tau$ & $5.71\times 10^{-5}$ & $4.38\times 10^{-8}$ \\
    
    \end{tabular}\\
\hline
\hline
\end{tabular}
\caption{Two sets of benchmark points for $1^{\rm st}$ and $2^{\rm nd}$ Generation singlet-doublet fermions are given. Where for the Set 1,  $M_{\phi_2}=532.14$ GeV and two non zero neutrino mass eigen values are $\sqrt{\Delta m^2_{\rm atm}}=4.94\times 10^{-11}$ GeV and $\sqrt{\Delta m^2_{\rm sol}}=8.63\times10^{-12}$ GeV. Similarly, for the Set 2, $M_{\phi_2}=711$ GeV, $\sqrt{\Delta m^2_{\rm atm}}=4.95\times 10^{-11}$ GeV and $\sqrt{\Delta m^2_{\rm sol}}=8.58\times10^{-12}$ GeV.}
\label{tab:benchmarks}
\end{center}
\end{table}
}}

Imposing the bound from cosmological data on the sum of light neutrino masses, $\sum m_i < 0.12$ eV and the neutrino oscillation data ($\Delta m^2_{\rm atm}=(2.358-2.544)\times10^{-3}~{\rm eV}^2$ and $\Delta m^2_{\rm sol.}=(6.79-8.01)\times10^{-5}~{\rm eV}^2$) at 3$\sigma$ C.L., we put stringent constraints on the parameters governing the generation of the light Dirac neutrino mass. Two benchmark values are given in Table \ref{tab:benchmarks}. For simplicity, we fixed ($M_{\phi_1}-M_{\phi_2}$) to be 1 GeV and the coupling parameters of the second generation singlet-doublet fermions comparably small to that of first generation.
It is noteworthy that this constraint not only establishes a distinctive correlation between the neutrino mass and the $(g-2)_\mu$, but also forges intriguing links between the DM phenomenology and the model's predictions regarding $\Delta N_{\rm eff}$. 
\begin{table}[h!]
\begin{center}
\begin{tabular}{||@{\hspace{0cm}}c@{\hspace{0cm}}||@{\hspace{0cm}}c@{\hspace{0cm}}||}
				\hline
				\hline
				\begin{tabular}{c}
					{\bf Model}\\
					{\bf Parameters}\\ 
					\hline
					$\sin\theta$ \\
    \hline
    $\Delta M=M_{\chi_1}-M_{\chi_2}$ (GeV) \\
    \hline
    $\Delta M_1=M_{\phi_2}-M_{\chi_2}$ (GeV) \\
    \hline
    $M_{\chi_2}$ (GeV)  \\
    \hline
    ${\lambda_\psi}_e$  \\
    \hline
    ${\lambda_\psi}_\mu$  \\ 
    \hline
    ${\lambda_\psi}_\tau$  \\ 
    \hline
    ${\lambda_\chi}_e$  \\ 
    \hline
    ${\lambda_\chi}_\mu$  \\ 
    \hline
    ${\lambda_\chi}_\tau$ \\ 
				\end{tabular}
				&
		      \begin{tabular}{c|c}
			\multicolumn{2}{c}{\bf Range for the scan}\\
			\hline
			$1^{\rm st}$ Generation&$2^{\rm nd}$ Generation \\
			\hline
			$[10^{-8}, 0.3]$ & $0.7$\\
            \hline
             $[1, 10^{3}]$ & $[500, 10^{3}]$ \\
            \hline
            $[1, 600]$ & $-$ \\
            \hline
            $[1, 10^{3}]$ & $[2000, 10^{4}]$ \\
             \hline
            $[10^{-8}, 10^{-5}]$ & $[10^{-1}, 10^{-7}]$ \\
             \hline
             $[1, 10^{-3}]$ & $[10^{-1}, 10^{-7}]$ \\ 
            \hline
          $[10^{-1}, 10^{-4}]$ & $[10^{-1}, 10^{-7}]$ \\ 
          \hline
             $[10^{-7}, \sqrt{4\pi}]$ & $[10^{-10}, 10^{-7}]$ \\ 
             \hline
              $[10^{-7}, \sqrt{4\pi}]$ & $[10^{-10}, 10^{-7}]$ \\ 
            \hline
             $[10^{-7}, \sqrt{4\pi}]$ & $[10^{-10}, 10^{-7}]$ \\ 
			\end{tabular}\\
			\hline
			\hline
		\end{tabular}
\caption{Range of model parameters used for the numerical scan. We fix the mass difference, $M_{\phi_1}-M_{\phi_2}=1~{\rm GeV}$.}
\label{tab:param}
\end{center}
\end{table}

	\section{Anomalous Magnetic Moment of Muon: $(g-2)_\mu$}
	\label{section:mug2}
	
	The anomalous magnetic moment of the muon, often denoted as $(g-2)_\mu$, refers to the difference between the muon's actual magnetic moment and the value predicted by the Dirac equation within the framework of quantum electrodynamics (QED). This deviation arises due to quantum fluctuations and interactions with virtual particles in the vacuum. 
	
	\begin{figure}[h]
		\centering
		\includegraphics[scale=0.1]{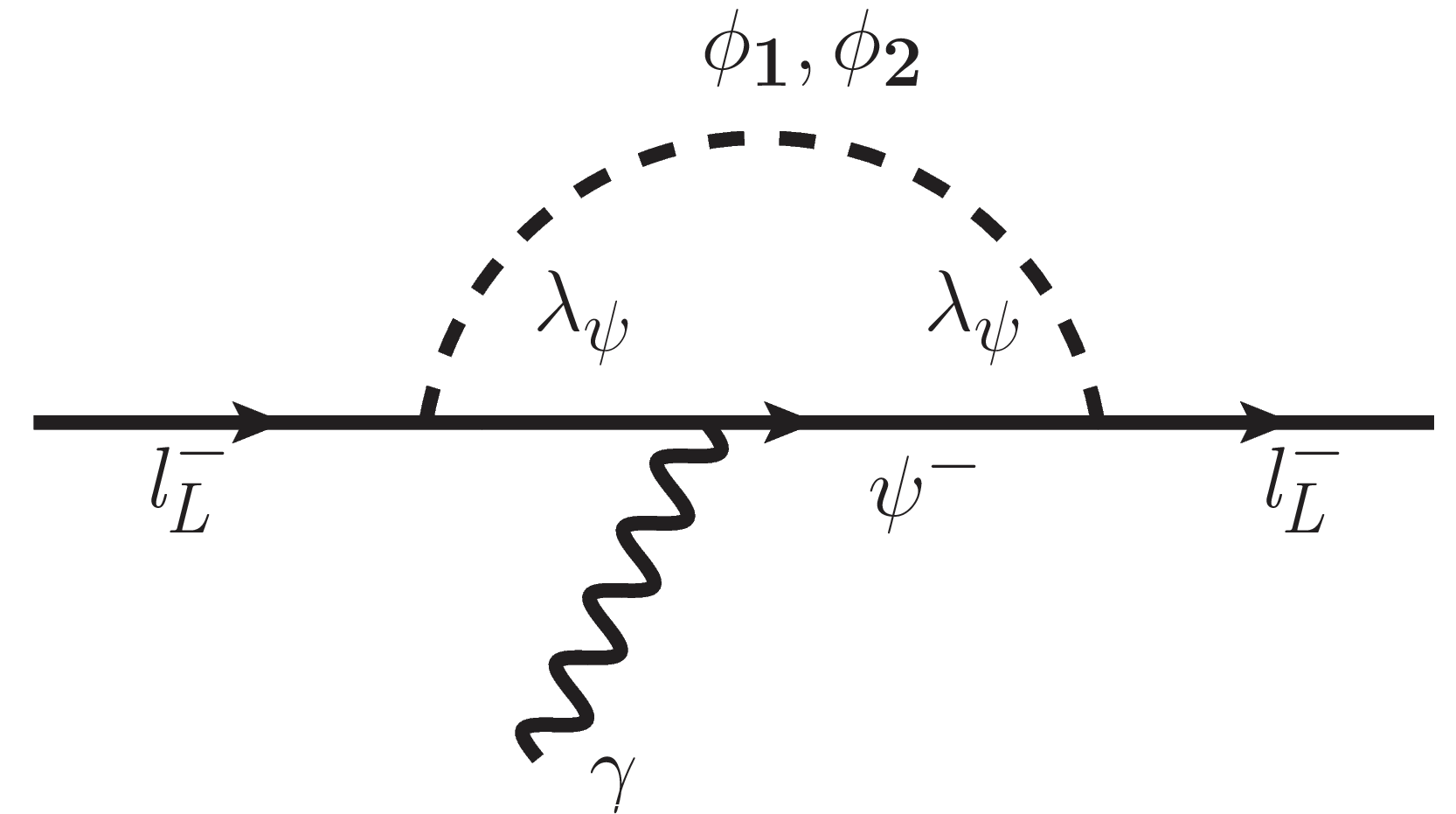}
		\caption{Feynman diagram giving rise to $(g-2)_\mu$ and lepton flavour violation.}
		\label{fig:g-2}
	\end{figure}

	In our setup, the new positive contribution to the $(g-2)_\mu$ comes from the one loop diagram with the charged doublet fermion $\psi^{-}$ and the singlet scalars $\phi_{1,2}$ in the loop. This contribution to $(g-2)_\mu$ is given by\cite{Lindner:2016bgg},
	
	\begin{equation}\label{eq:g2}
    \Delta a_{\mu}=\frac{m_{\mu }^2 \left(y_{\psi }\right)_{\mu }^2}{8 \pi ^2} \int_0^1 dx \left[I_1^{+}(x,M_{\psi^-},M_{\phi_1})-I_1^{-}(x,M_{\psi^-},M_{\phi_2})\right]
\end{equation}
Where
\begin{equation}
    I_1^{(\pm)}(x,m_1,m_2)=\frac{x^2 \left(1-x\pm\frac{m_{1}}{m_{\mu }}\right)}{\left((1-x) \left(m_{2}^2 -x m_{\mu }^2\right)+x m_1^2\right)}
\end{equation}
	
\begin{figure}[h]
		\centering
		\includegraphics[scale=0.45]{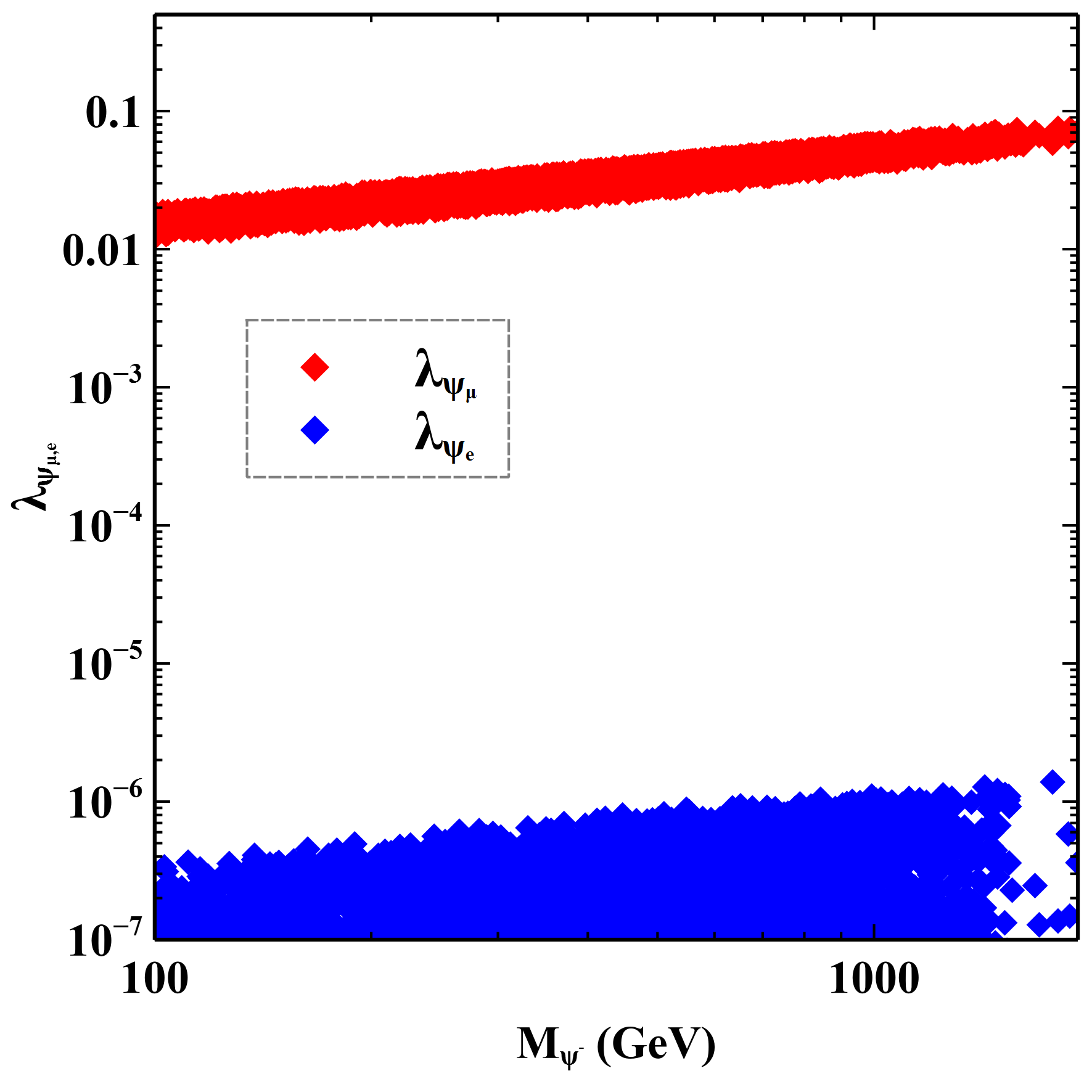}
		\caption{The allowed region for $(g-2)_\mu$ is depicted by the red coloured points in the parameter space of $\lambda_{\psi_\mu}$ and $M_{\psi^-}$. The values of $\lambda_{\psi_e}$ consistent with the LFV constraints are also projected in the same plane by the blue points. The details of the LFV decay are discussed in the next section.}
		\label{fig:g2lfv1}
	\end{figure}

In Fig.~\ref{fig:g2lfv1}, we showcase the parameter space satisfying correct $\Delta a_\mu$ by the red coloured points, in the plane of $\lambda_{\Psi_\mu}$ and $M_{\psi^-}$ with $M_{\phi_1}$ varying in a range as mentioned in Table~\ref{tab:param}. We first carried out a numerical scan varying the model parameters in the range as shown in Table~\ref{tab:param} and found the parameter space that satisfies the correct neutrino mass criteria. The same parameter set is then used for the scan to check the constraint from $(g-2)_\mu$. 
	
	Clearly, we can see that $\lambda_{\Psi_\mu}$ is constrained in the interval $[0.01,0.1]$ while imposing the lower bound on the charged fermion doublet mass to be $M_{\Psi^{-}} > 102.7$ GeV from LEP~\cite{Barbieri:2004qk}. As the mass of the particles in the loop suppresses the contribution to $(g-2)_\mu$, the corresponding coupling $\lambda_{\psi_\mu}$ has to increase. This can be easily seen from Fig. \ref{fig:g2lfv1} 
	
	\subsection{Lepton Flavour Violation}
	The decay involving charged lepton flavour violation (LFV) is a significant process for investigating physics beyond the SM. Within the SM framework, this process takes place at the one-loop level and is greatly suppressed due to the minuteness of neutrino masses and therefore remains far beyond the current experimental sensitivities ~\cite{MEG:2016leq}. Consequently, any future detection of LFV decays, such as $\mu\to e\gamma$, would unquestionably constitute evidence of physics beyond the SM. In our current model, this additional new physics contribution comes from the same diagram as shown in Fig.~\ref{fig:g-2} with leptons of two different flavours in the external legs. The branching ratio for $\mu \to e \gamma$ is given by \cite{Lindner:2016bgg}:

	\begin{equation}\label{eq:lfv}
	{\rm BR}(\mu^- \rightarrow e^- \gamma)=\frac{3 \alpha}{16\pi G_F^2} \left|\lambda_{\psi_\mu} \lambda_{\psi_e}I(M_{\psi^-}, M_\phi)\right|^2,
	\end{equation}
	where
	\begin{equation*}
	I(M_{\psi^-}, M_\phi)=\int^1_0 dx \int^{1-x}_0 dy ~\left[I_2^+ (x, y, M_{\psi^-}, M_{\phi_1})-I_2^- (x, y, M_{\psi^-}, M_{\phi_2})\right]
	\end{equation*}
	and
	\begin{equation*}
	I_2^{(\pm)} (x, y, m_1, m_2)=\frac{x\left(y+(1-x-y)\frac{m_e}{m_\mu}\right)\pm(1-x)\frac{m_1}{m_\mu}}{-xym_\mu^2 - x (1 - x - y)m_e^2 + xm_2^2 + (1 - x)m_1^2}.
	\end{equation*}

Given the existing constraints on $\lambda_{\psi_\mu}$ and $M_{\psi^{-},\phi_1}$ from the $(g-2)_\mu$ experiment, and taking into account the upper limit on $Br(\mu\to e\gamma)$ set by MEG~\cite{MEG:2016leq}, we can establish an upper bound for $\lambda_{\psi_e}$. This is shown in Fig. \ref{fig:g2lfv1} by the blue coloured points. As we can see, it puts a conservative upper limit that $\lambda_{\psi_e}$ must be less than $10^{-6}$ for $M_{\psi^-} \in [100, 1800]~{\rm GeV}$ in order to remain consistent with the MEG constraint.

\section{Dark Matter Phenomenology}
\label{section:DMpheno}
As outlined in section~\ref{section:model}, $\chi_2$, being the lightest odd particle under $Z_2$ symmetry, attains stability and emerges as a plausible DM candidate in our framework. Being an admixed state of a singlet and a doublet fermion, its thermalization and relic abundance are crucially governed by the Yukawa and gauge interactions. Thus its relic density is determined by the process of thermal freeze-out, wherein the primary mechanisms involve the annihilation of DM into SM particles and $\nu_{R}$, along with co-annihilation among the dark sector constituents $\chi_2$, $\chi_1$, $\phi_1$, $\phi_2$,and $\psi^{-}$. 
	
Before delving into the phenomenology of DM, it is important to note that the inclusion of $\nu_{R}$ and $\phi_{1,2}$ in this setup, and their significant involvement in the genesis of neutrino mass at one-loop, not only offers an additional cosmological probe to validate the model but also alters the conventional outcomes typically observed in the study of Dirac fermionic singlet-doublet DM alone as explored in~\cite{Bhattacharya:2018fus, Bhattacharya:2017sml}. As in section~\ref{section:delneff}, we will thoroughly examine the $\Delta N_{\rm eff}$ component of this scenario, it is pivotal to consider whether $\nu_{R}$ undergoes thermalization or not because the same coupling plays a crucial role in determining DM relic density. 
	
The production of RHNs ($\nu_{R_i}$) primarily depends on the coupling parameter $\lambda_\chi$. This same coupling also dictates whether RHNs undergo thermalization. For values of $\lambda_\chi$ greater than $10^{-3}$, we observe that RHNs were in thermal equilibrium with the SM bath alongside other constituents of the dark sector. Consequently, both $\phi_{1,2}$ and $\chi_{1,2}$ annihilation contribute to the production of RHNs and hence the enhancement of additional relativistic energy density. Conversely, for smaller values of $\lambda_\chi$, RHNs were not in equilibrium with the SM bath, making it challenging to produce them through thermal processes. In such instances, RHNs can still be generated through non-thermal freeze-in processes, particularly via the decay of $\phi_{1,2}$ in the process $\phi_{1,2}\rightarrow\chi_2 \nu_{R}$.
	
	Consequently, we categorize our analysis into two distinct cases:
	
	\noindent(i) {\bf Case-1}: $10^{-3}<\lambda_\chi <\sqrt{4\pi} $.\\
	(ii) {\bf Case-2}: $10^{-7}<\lambda_\chi \leq 10^{-3} $.\\
where we use $\lambda_\chi=(\lambda_{\chi_e}+\lambda_{\chi_\mu}+\lambda_{\chi_\tau})/3$.

	\subsection{Relic Abundance of DM}
	\label{subsec:relic}
	The relic density of DM in this scenario is achieved by solving the Boltzmann equation 
	\begin{eqnarray}
	\frac{dn}{dt} + 3 H n = - \langle \sigma v \rangle_{\rm eff} (n^{2} - (n^{\rm eq})^{2})
	\end{eqnarray}
	where $n=\sum_i n_i $ represents the total number density of all the dark sector particles and  $n^{\rm eq}$ is the equilibrium number density.  $\langle \sigma v \rangle_{\rm eff}$ represents the effective annihilation cross-section which takes into account all number
	changing processes for DM freeze-out is given by:
 
		\begin{eqnarray}
		\label{eq:effcrs}
		\langle \sigma v \rangle_{\rm eff} &=& \frac{g^2_2}{g^2_{\rm eff}}\langle\sigma v\rangle_{\chi_{_2}\chi_{_2}}+\frac{g_2 g_{\phi_{1,2}}}{g^2_{\rm eff}}\langle\sigma v\rangle_{\chi_{_2}{\phi_{1,2}}} (1+\Delta_{\phi_{1,2}})^{{3}/{2}}\exp(-x\Delta_{\phi_{1,2}})  \nonumber \\ &{}& + \frac{g_2 g_1}{g^2_{\rm eff}}\langle\sigma v\rangle_{\chi_{_2}\chi_{_1}} (1+\Delta_{\chi{_1}})^{{3}/{2}}\exp(-x\Delta_{\chi_{_1}}) \nonumber\\&{}&+\frac{g_2 g_3}{g^2_{\rm eff}}\langle\sigma v\rangle_{\chi_{_2}\psi^{-}} (1+\Delta_{\psi^{-}})^{{3}/{2}}\exp(-x\Delta_{\psi^{-}})+\frac{g^2_{\phi_{1,2}}}{g^2_{\rm eff}}\langle\sigma v\rangle_{{\phi_{1,2}}{\phi_{1,2}}} (1+\Delta_{\phi_{1,2}})^{3}\exp(-2x\Delta_{\phi_{1,2}})\nonumber\\&{}&+\frac{g_{\phi_{1,2}} g_{1}}{g^2_{\rm eff}}\langle\sigma v\rangle_{{\phi_{1,2}}\chi_{1}} (1+\Delta_{\phi_{1,2}})^{3/2}(1+\Delta_{\chi_1})^{3/2}\exp\left(-x(\Delta_{\phi_{1,2}}+\Delta_{\chi_1})\right)\nonumber\\&{}&+ \frac{g_{\phi_{1,2}} g_{3}}{g^2_{\rm eff}}\langle\sigma v\rangle_{{\phi_{1,2}}\psi^{-}} (1+\Delta_{\phi_{1,2}})^{3/2}(1+\Delta_{\psi^{-}})^{3/2}\exp\left(-x(\Delta_{\phi_{1,2}}+\Delta_{\psi^{-}})\right)\nonumber\\&{}&+\frac{g^2_1}{g^2_{\rm eff}}\langle\sigma v\rangle_{\chi_{_1}\chi_{_1}} (1+\Delta_{\chi_1})^{3}\exp(-2x\Delta_{\chi_1})+\frac{g^2_3}{g^2_{\rm eff}}\langle\sigma v\rangle_{\psi^{+}\psi^{-}} (1+\Delta_{\psi^{-}})^{3}\exp(-2x\Delta_{\psi^{-}}) \nonumber\\&{}&+\frac{g_1 g_{3}}{g^2_{\rm eff}}\langle\sigma v\rangle_{\chi_1 \psi^{-}} (1+\Delta_{\chi_1})^{3/2}(1+\Delta_{\psi^{-}})^{3/2}\exp(-x(\Delta_{\chi_1}+\Delta_{\psi^{-}}))
	\end{eqnarray} 
where $g_1, g_2, g_3, g_{\phi_{1}} ~{\rm and}~g_{\phi_{2}}$ represent the internal degrees of $\chi_1,\chi_2,\psi^{-}$, ${\phi_{1}}$ and ${\phi_{2}}$ respectively and $\Delta_i$ stands for the ratio $(M_i-M_{\chi_2})/M_{\chi_{2}}$ with $M_i$ denoting the mass of $\chi_1,\psi^{-},{\phi_{1}}~{\rm and}~{\phi_{2}}$. Here $g_{\rm eff}$ is the effective degree of freedom which is given by:
\begin{eqnarray}
g_{\rm eff}&=& g_2 + g_{\phi_{1,2}} (1+\Delta_{\phi_{1,2}})^{3/2}\exp(-x\Delta_{\phi_{1,2}})\nonumber\\&+& g_1 (1+\Delta_{\chi_1})^{3/2}\exp(-x\Delta_{\chi_1})\nonumber\\&+& g_3 (1+\Delta_{\psi^{-}})^{3/2}\exp(-x\Delta_{\psi^{-}})
\end{eqnarray}
and $x$ is the dimensionless parameter $M_{\rm \chi_2}/T$.

	The relic density of DM $\chi_2$ can then be evaluated as :
	\begin{eqnarray}
	\Omega_{\chi_2} h^2 = \frac{1.09 \times 10^9 {\rm GeV}^{-1}}{\sqrt{g_* }M_{Pl}} \left[\int_{x_f}^\infty dx~\frac{\langle \sigma v \rangle_{\rm eff}}{x^2}\right]^{-1}
	\end{eqnarray}
	Here $x_f =M_{\chi_2}/T_{f}$, and $T_f$ denotes the freeze-out temperature of $\chi_2$.
	
To gain insight into the relic density of DM and the specific influence of model parameters in achieving the observed relic density, we conducted various analyses and explored the allowed parameter space. As studied in~\cite{Bhattacharya:2018fus, Borah:2021khc}, the pivotal parameters governing the relic abundance of a singlet-doublet Dirac fermionic DM include: the mass of the DM ($M_{\chi_2}$), the mass splitting between the dominant singlet and dominant doublet physical states ($\Delta M$), and the mixing angle ($\sin\theta$). In addition to these three parameters, in the present setup, we have additional parameters that also affect the relic density of DM because of the presence of $\phi_{1,2}$ and $\nu_R$. The presence of $\phi_{1,2}$ in the dark sector leads to additional co-annihilation processes whereas $\nu_{R}$ facilitates a new annihilation channel of DM to RHN. Thus the mass difference between $\phi_2$ and $\chi_2$, ($\Delta M_{1} = M_{\phi_2} - M_{\chi_2}$), the scalar quartic coupling, $\lambda_{\phi H}$ as well as $\lambda_\chi$ (DM- $\nu_R-\phi_{1,2}$ coupling)  also become important parameters in determining the relic density of DM. We have mentioned earlier that the mass difference between physical scalars ($M_{\phi_1}-M_{\phi_2}$) are kept fixed to 1 GeV. So, in this model, we have a multi-dimensional parameter space that decides the relic density of DM. It is worth mentioning here that we used the package MicrOmegas~\cite{Belanger:2014vza} for computing
	annihilation cross-sections and relic density, after generating the model files using LanHEP~\cite{Semenov:2008jy}.
	
	\begin{figure}[h]
		\centering
		\includegraphics[scale=0.45]{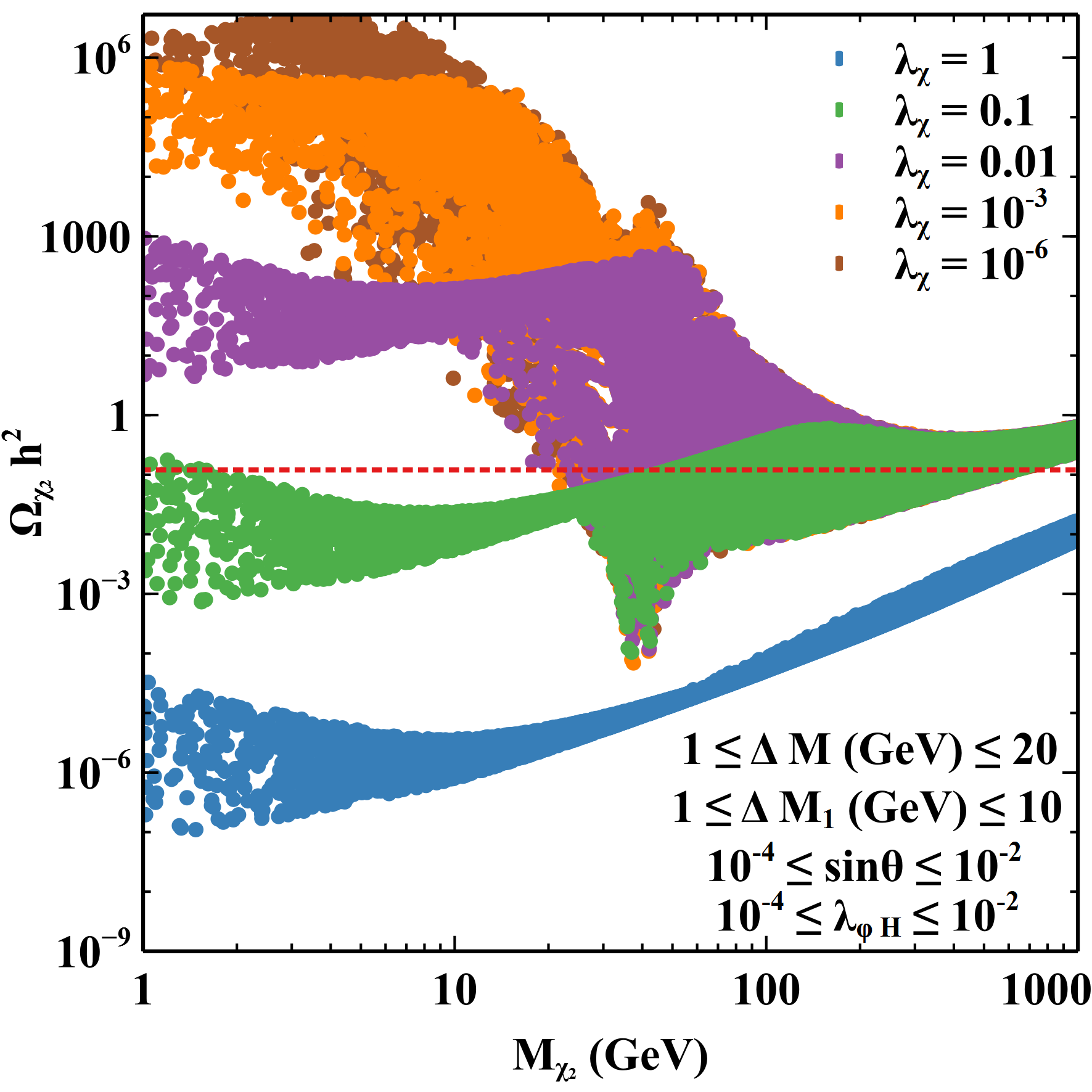}
		\caption{Relic density of DM as a function of DM mass for different values of $\lambda_\chi$, other parameters being varied randomly as mentioned in the inset.}
		\label{fig:relicplot1}
	\end{figure}

	In Fig.~\ref{fig:relicplot1}, we present the relic density of DM as a function of its mass, considering different benchmark values of $\lambda_\chi$ while randomly varying the other parameters $\Delta M$, $\Delta M_1$, $\sin \theta$, and $\lambda_{\phi H}$ as shown in the inset of the figure. It is evident that an increase in $\lambda_\chi$ results in a gradual decrease in the relic density. This is attributed to the fact that as $\lambda_\chi$ increases, the annihilation cross-section of $\chi_2$ into $\nu_R$ also increases. Consequently, the overall effective annihilation cross-section is enhanced, leading to a reduction in the relic density.
	Additionally, an interesting feature observed in Fig. 4 is that when $\lambda_\chi$ is not significantly large (i.e., $\lambda_\chi < 0.1$), the relic density of DM is influenced not only by its annihilation into $\nu_R$, but also by other annihilation and co-annihilation processes involving SM particles mediated by the gauge bosons. This explains the resonance features observed around $M_{V}/2~(V\equiv W^{\pm},Z)$, represented by the green, purple, orange, and brown points.
	The Higgs resonance is not prominently visible in this plot because the Higgs-mediated annihilation or co-annihilation channels are less efficient for small $\sin\theta$ and small mass-splitting $\Delta M$. This is attributed to the fact that the corresponding Yukawa coupling $y$ is proportional to $\Delta M \times \sin2\theta$, as illustrated in Eq.~\eqref{eq:mixang}.
	However, in cases where $\lambda_\chi$ significantly surpasses other couplings, the channel $\chi_2 \chi_2 \to \nu_R \nu_R$ becomes the predominant contributor to the relic density of $\chi_2$. Since this annihilation process occurs through a $t$-channel, resonance effects are not observed, as indicated by the blue points.
	
	
	\begin{figure}[h]
		\centering
		\includegraphics[scale=0.45]{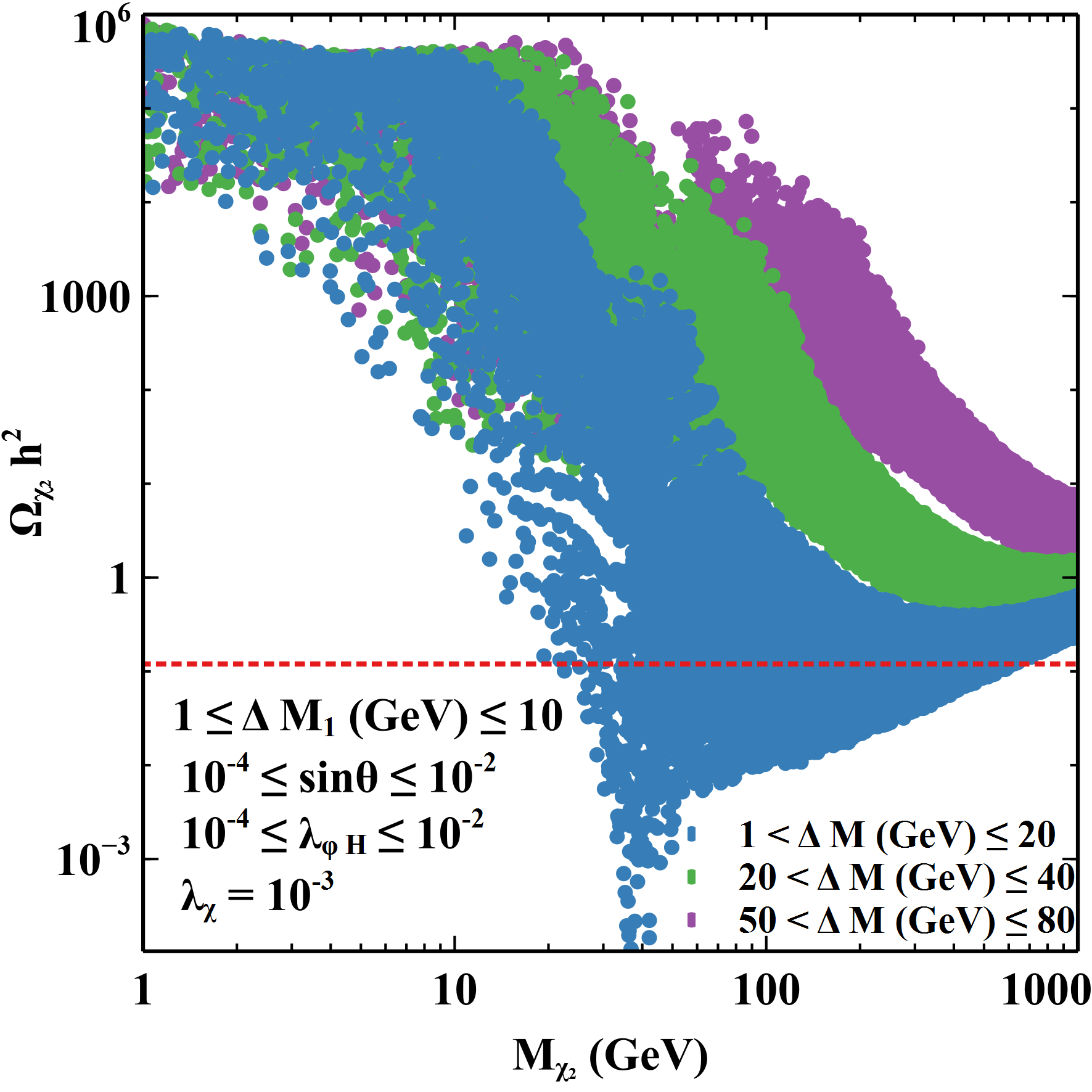}
		\caption{Relic density of DM as a function of DM mass for different values of $\Delta M$, other parameters being varied randomly as mentioned in the inset.}
		\label{fig:relicplot2}
	\end{figure}
	
	Fig. \ref{fig:relicplot2} illustrates the correlation between relic density and singlet and doublet mass splitting ($\Delta M$), while mitigating the influence of DM annihilation to $\nu_R$s by fixing $\lambda_\chi=10^{-3}$. Other relevant parameters are varied randomly in a specific range as mentioned in the inset. With such a choice of parameters, DM annihilation to $\nu_R$ remains subdominant while the DM annihilation to SM particles and co-annihilation of DM with $\chi_1$ and $\psi^-$ dominantly decide the relic abundance. As we can see, with an increase in the mass-splitting $\Delta M$, the relic density decreases and vice versa. This can be understood from Eq.~\eqref{eq:effcrs} which is the effective annihilation cross-section of DM. As $\langle\sigma v \rangle_{\rm eff}$ decreases with an increase in $\Delta_{\chi_1} = \Delta M / M_{\chi_2}$ due to exponential suppression, which consequently elevates the relic density. It is important to note that $\Delta_{\psi^-}$ is not an independent parameter but rather is dependent on $\Delta M$ and $\sin\theta$. 
	
	
	\begin{figure}[h]
		\centering
		\includegraphics[scale=0.45]{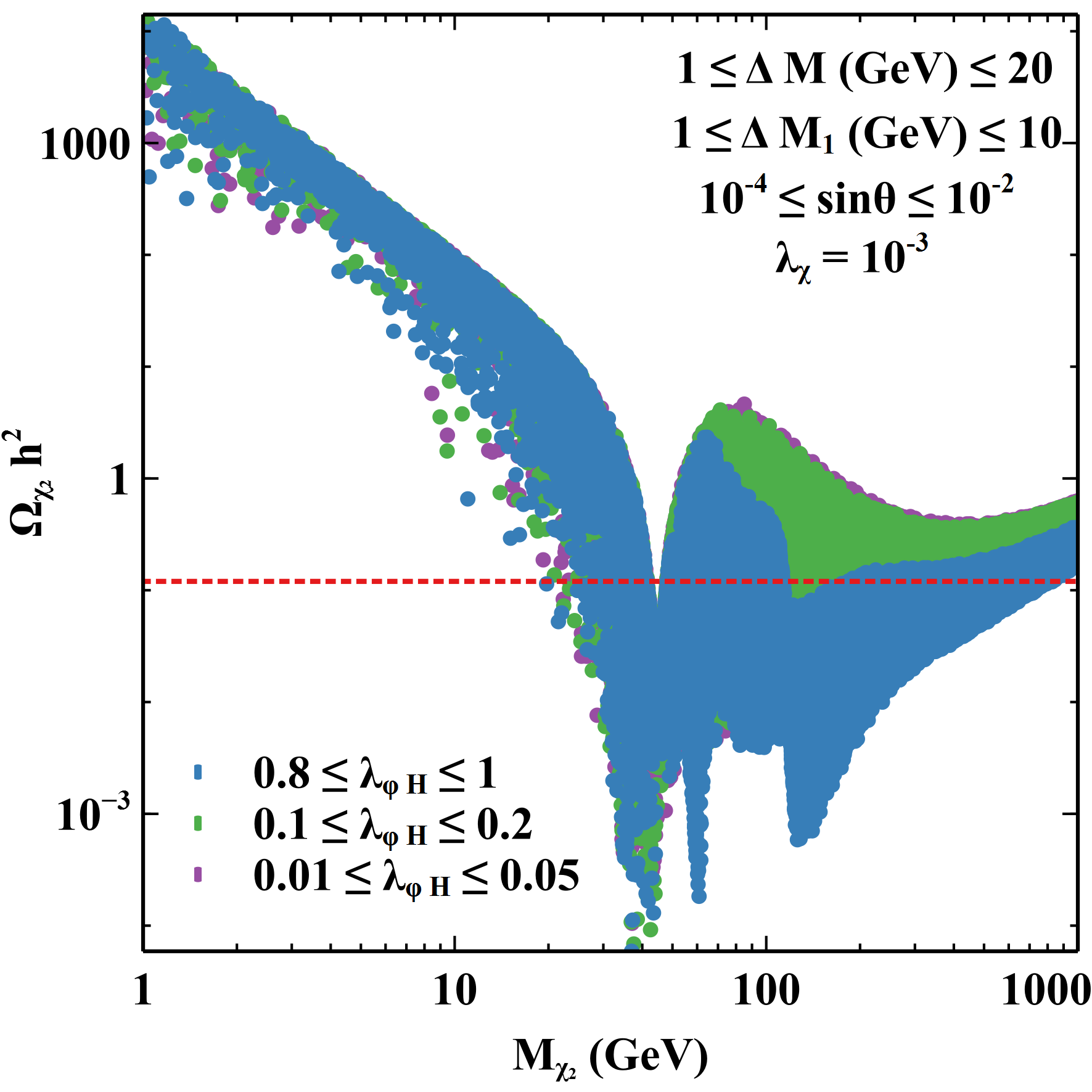}
		\caption{Relic density as a function of DM mass for different ranges of $\lambda_{\phi H}$.}
		\label{fig:relicplot3}
	\end{figure}
	
In Fig.~\ref{fig:relicplot3}, we depict $\Omega_{\chi_2} h^2$ as a function of $M_{\chi_2}$ for various selections of $\lambda_{H\phi}$ to demonstrate the dependency of relic density on this scalar quartic coupling. It is important to note that $\lambda_{H\phi}$ primarily affects the co-annihilation contribution in the effective annihilation cross-section, as it governs the rate of annihilation of $\phi_{1,2}$ as well as co-annihilation of $\phi_{1,2}$ and $\chi_2$ to SM particles. For smaller values of $\lambda_{H\phi}$, we observe that the relic density is not very sensitive. It is only when $\lambda_{\phi H}\sim 1$ that this co-annihilation contribution becomes significant, ultimately influencing the relic density, as indicated by the blue points. The dips in the relic density correspond to resonances corresponding to gauge bosons and the SM Higgs boson.
	
A common feature observed in all three figures is that in the higher DM mass region, away from resonances, an increase in DM mass leads to a gradual rise in relic density. This phenomenon arises from the fact that in the non-relativistic limit, the effective annihilation cross-section of DM is inversely proportional to its mass. Consequently, as the DM mass increases, $\langle \sigma v \rangle_{\rm eff}$ decreases, resulting in an increase in relic density.
	
\subsection{Direct Detection}
\label{subsec:dd}
In this scenario, owing to the singlet-doublet mixing, the DM $\chi_2$ can interact with the target nucleus in terrestrial direct search experiments through $Z$ and Higgs mediated processes.  This leads to the cross-section for $Z$-boson mediated DM-nucleon scattering to be~\cite{ Goodman:1984dc,Essig:2007az}:
\begin{equation}\label{DDZ}
	\sigma_{\rm SI}^Z = \frac{G^2_F \sin^4\theta}{\pi A^2 }\mu_r^2 \Big|\left[ Z f_p + (A-Z)f_n \right]^2\Big|^2   
\end{equation}
Where the $f_p=f_n=0.33$ corresponds to the form factors for proton and neutron, respectively. Here $\mu_r$ is the reduced mass of the DM-nucleon system and  $A, ~Z$ are mass number and atomic number respectively. Similarly, the spin-independent 
DM-nucleon scattering cross-section through Higgs mediation is given by:
	\begin{equation}
	\label{DDH}
	\begin{aligned}
	\sigma^h_{\rm SI} &= \frac{4}{\pi A^2}\mu^2_r\frac{y^2 \sin^2 2\theta}{M^4_h}\Big[\frac{m_p}{v}\Big(f^{p}_{Tu} + f^{p}_{Td} + f^{p}_{Ts} + \frac{2}{9}f^{p}_{TG}\\
	&+\frac{m_n}{v}\Big(f^{n}_{Tu} + f^{n}_{Td} + f^{n}_{Ts} + \frac{2}{9}f^{n}_{TG}\Big)\Big]^2
	\end{aligned}
	\end{equation}
	Different coupling strengths between DM and light quarks are
	given by \cite{Bertone:2004pz,Alarcon:2012nr} $f^p_{Tu}=0.020 \pm 0.004,~f^p_{Td}=0.026 \pm
	0.005,~f^p_{Ts}=0.014 \pm 0.062,~ f^n_{Tu}=0.020 \pm 0.004,~ f^n_{Td}=0.036 \pm 0.005, ~f^n_{Ts}=0.118 \pm 0.062$. The coupling of DM with the gluons
	in target nuclei are parameterized by \cite{Hoferichter:2017olk} $f^{p,n}_{TG}=1-\sum_{q=u,d,s}f^{p,n}_{Tq}$.
	
	\begin{figure}[h!]
		\centering
		\includegraphics[scale=0.1]{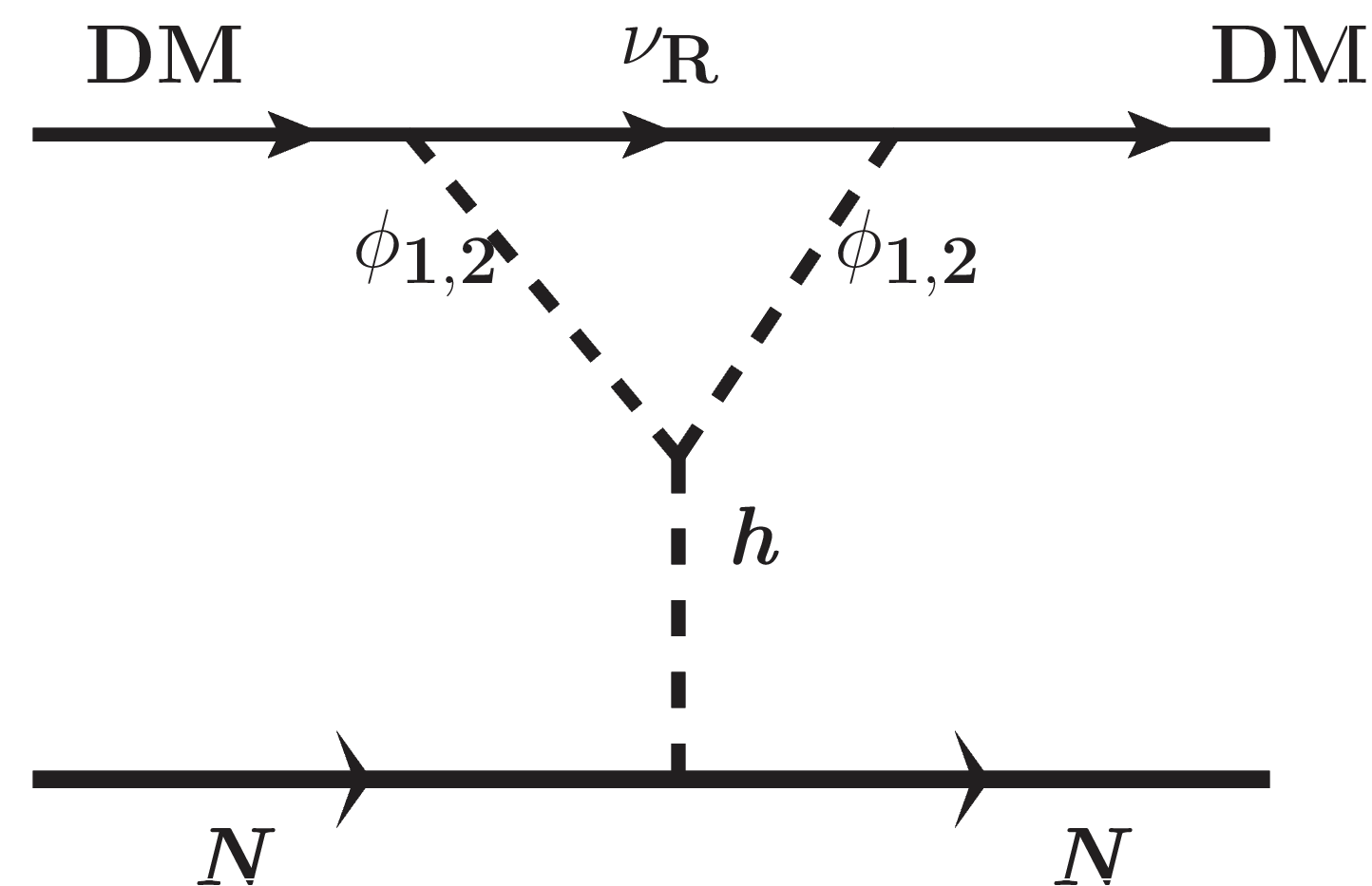}
        \includegraphics[scale=0.1]{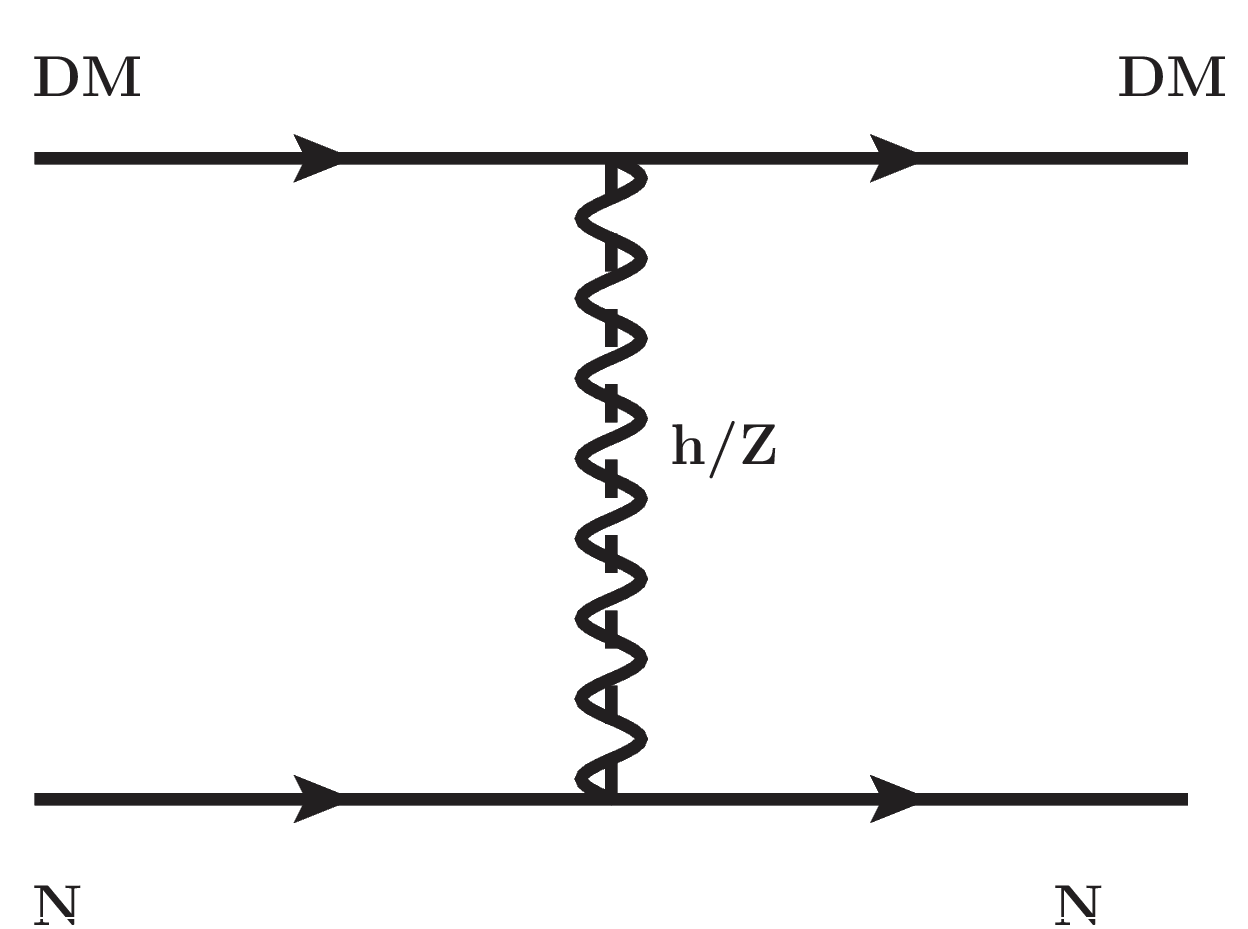}
		\caption{ Spin-independent elastic DM-nucleon scattering
			arising at one loop and tree level.}
		\label{fig:ddfeynmann}
	\end{figure}

	\begin{figure}[h]
		\centering
		\includegraphics[scale=0.4]{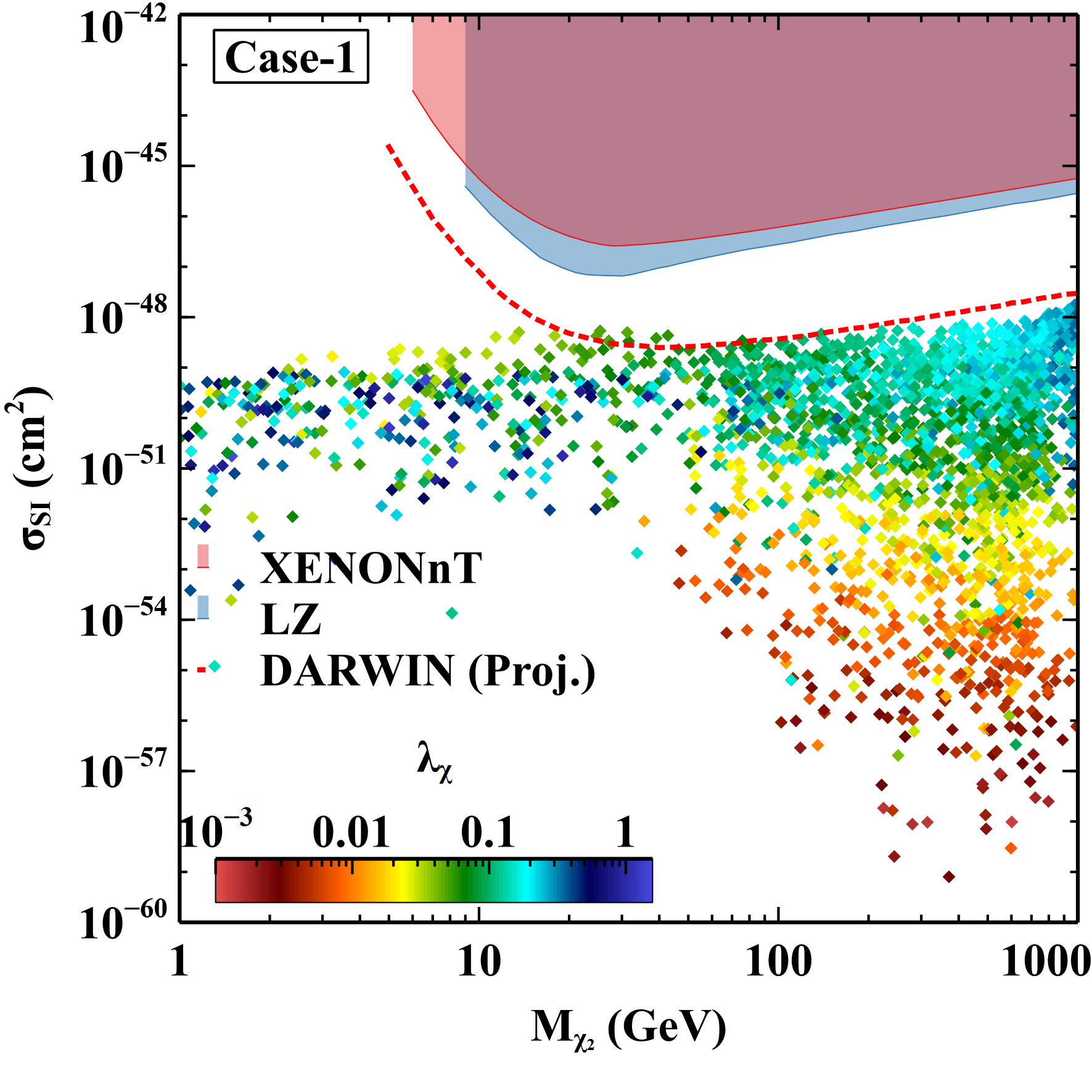}
		\hfil
		\includegraphics[scale=0.4]{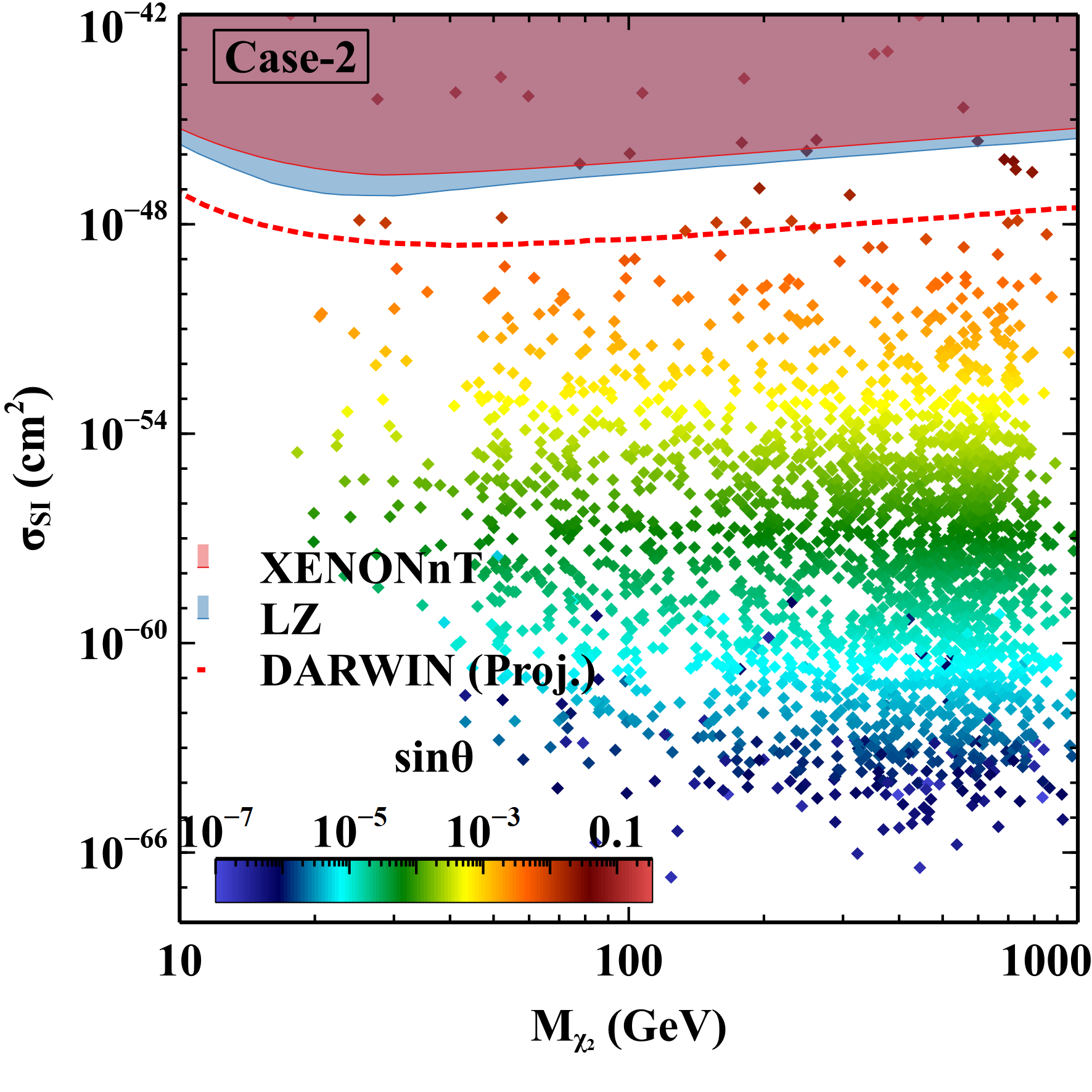}
		\caption{Spin-independent DM-nucleon scattering cross-section as a function of DM mass for case-1(left) and case-2(right). }
		\label{fig:dd2}
	\end{figure}
	
Besides these tree-level processes for DM nucleon scattering, there exists another contribution to the spin-independent direct search cross-section, which arises at the loop level with $\nu_R$ and $\phi_{1,2}$ in the loop. The corresponding Feynman diagram is as shown in Fig.~\ref{fig:ddfeynmann}.

	This cross-section can be evaluated as~\cite{Ibarra:2016dlb}:
	\begin{equation}\label{eq:dd1}
	\sigma^{\rm loop}_{\rm SI}=\frac{4}{\pi}\frac{M^2_{\chi_2} m_p^2}{(M_{\chi_2}+m_p)^2}m_p^2~\mathcal{C}^2~f_q^2
	\end{equation}
	Where the form factor, $f_q\simeq0.3$ and the loop-induced effective coupling $\mathcal{C}$ is given by,
	\begin{equation*}
	\mathcal{C} = \frac{\lambda_\chi^2}{16\pi^2 M_h^2M_{\chi_2}}\lambda_{\phi H}G\left(\frac{M_{\chi_2}^2}{M_\phi^2}\right)
	\end{equation*}
	And the loop function $G(x)$ is given by,
	\begin{equation*}
	G(x) = \frac{x+(1-x){\rm ln}(1-x)}{x}
	\end{equation*}
	
	From Eqs.~\eqref{DDZ},\eqref{DDH} and \eqref{eq:dd1}, it is clear that the DM-nucleon interaction strength crucially depends on $\sin\theta$ at tree level, and $\lambda_\chi$ and $\lambda_{\phi H}$ at loop level. 
	Here, it is worth noting that, as $\lambda_\chi$ and $\sin\theta$ are correlated through the neutrino mass constraint as discussed in section~\ref{section:numass}, depending on their relative strength, the tree level or loop level contribution to DM-nucleon scattering dominates. We scrutinize the model parameters in both cases as mentioned at the beginning of section \ref{section:DMpheno} against the most stringent constraint on DM-nucleon scattering cross-section from XENON-nT~\cite{XENON:2023cxc} and LZ~\cite{LZ:2022lsv} experiments. We also showcase the projected sensitivity of the DARWIN ~\cite{DARWIN:2016hyl}. In case-1 ($\lambda_{\chi} > 10^{-3}$), because of loop-suppression $\sigma^{\rm loop}_{\rm SI}$, as well as because of very small $\sin\theta$, $\sigma^{Z,h}_{\rm SI}$, remains far below the present sensitivity of LZ and Xenon-nT and hence most of the parameter space remains safe from the DM direct search bounds. This is visible from the left panel of Fig.~\ref{fig:dd2}. In case-2($\lambda_\chi<10^{-3}$), the one-loop contribution remains suppressed throughout and only the tree-level diagrams contribute to the DM-nucleon scattering. The right panel of Fig.~\ref{fig:dd2} clearly illustrates that when $\sin\theta$ is substantial, the interaction strength is correspondingly high, resulting in a large DM-nucleon cross-section. As a result, direct search experiments impose significant constraints on $\sin\theta$. This stringent upper limit is $\sin\theta<0.03$.

After incorporating the constraints from relic abundance and direct search of DM, now we show the final parameter space in the plane of DM mass and the mass-splittings $\Delta M$ and $\Delta M_1$ for the two different cases (Case-1 [$10^{-3}<\lambda_\chi <\sqrt{4\pi}$] and Case-2[ $10^{-11}<\lambda_\chi \leq 10^{-3} $] ) as mentioned earlier. We also impose the constraints from LEP experiment on the mass of the charged fermion doublet($M_{\psi^-}$) {\it i.e.} $M_{\psi^-}>102.7$ GeV and restrict the mass of $\phi_2$ to be greater than $M_{H}/2$ to avoid the constraints from Higgs invisible decay.

\vspace{0.3cm}
	
\noindent\textbf{Case-1 [$10^{-3}<\lambda_\chi <\sqrt{4\pi}$] :}
	
\begin{figure}[h]
		\centering
		\includegraphics[scale=0.4]{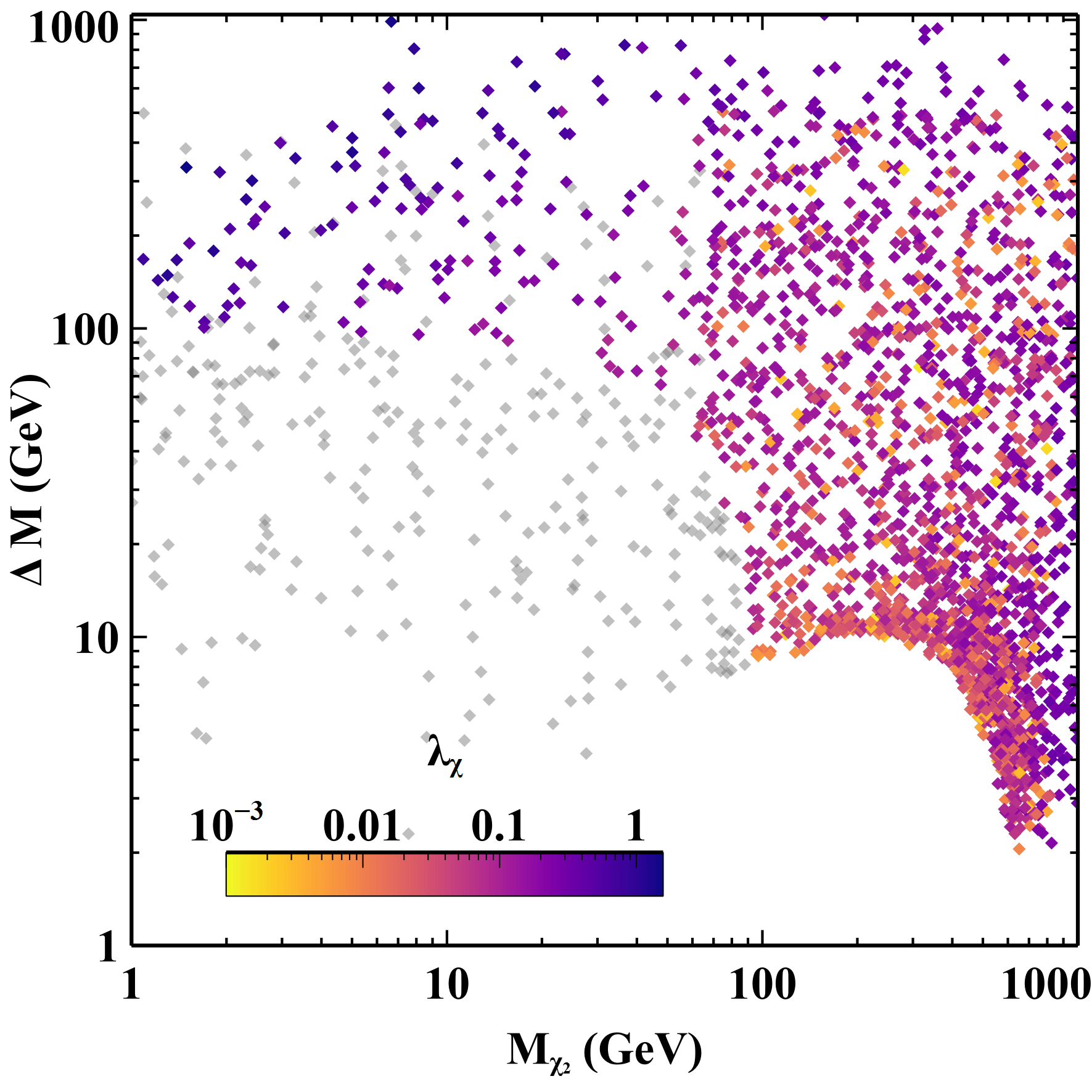}
		\includegraphics[scale=0.4]{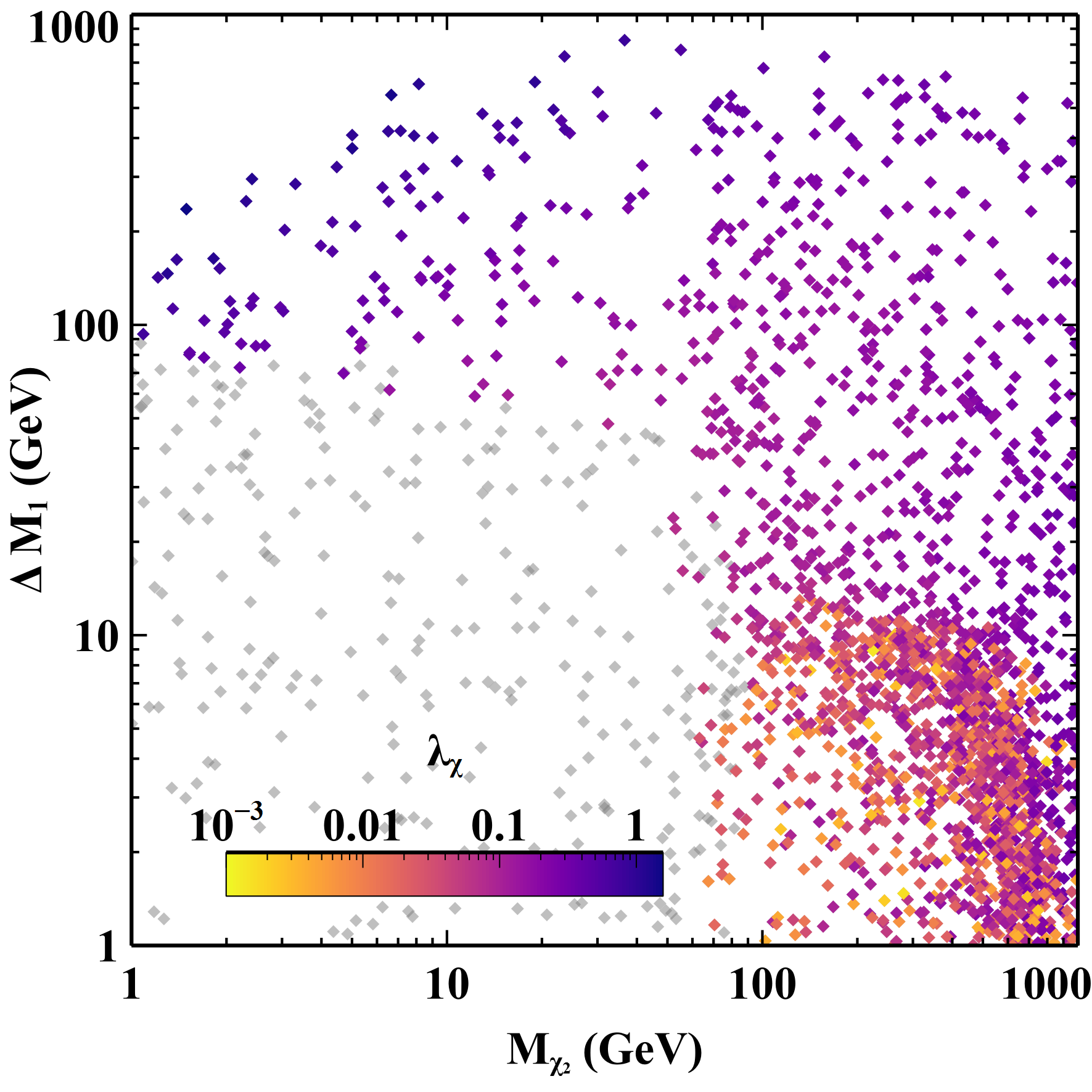}
		\caption{ Correct relic density and direct detection constraint satisfying points in the plane of $M_{\chi_2}$ and $\Delta M$ ([Left]) and in the plane of $M_{\chi_2}$ and $\Delta M_1$([Right]). The color code depicts the value of  $\lambda_\chi$ . Grey regions are ruled out by the LEP constraint.}
		\label{fig:c1relic1}
	\end{figure}

As we know in case-1, characterized by a large value of $\lambda_\chi$, the primary process determining the correct relic density of $\chi_2$ is $\chi_2 \chi_2 \to \nu_R \nu_R$ annihilation. Nevertheless, the contributions from co-annihilation, involving both $\phi_{1,2}$ annihilation to the SM and co-annihilation involving $\chi_1$ and $\psi^{-}$, also play crucial roles. These contributions depend on the mass splittings $\Delta M_1$ and $\Delta M$.
	From the left panel of Fig.~\ref{fig:c1relic1}, it is evident that as the mass splitting $\Delta M$ increases for DM mass $M_{\chi_2}\lsim 100$ GeV, $\lambda_\chi$ must also be raised to attain the correct relic density. This adjustment is necessary because, with an increase in $\Delta M$, the co-annihilation effect gradually diminishes, consequently reducing the effective cross section $\langle \sigma v \rangle_{\rm eff}$. Therefore, an increase in $\lambda_\chi$ is required to offset this effect and achieve the correct relic density. In the region where $M_{\chi_2}\lsim 100$ GeV and $\Delta M\lsim 100$ GeV, most of the points shaded in grey are excluded due to the imposed constraint on $M_{\psi^-}$ from the LEP experiment.
	As we increase the DM mass($M_{\chi_2} \gsim 100$ GeV), we observe that in the lower mass-splitting region ($\Delta M \lsim 15$ GeV), the mass splitting gradually decreases to achieve the correct relic density. This is because as the DM mass increases, the effective cross section $\langle \sigma v \rangle_{\rm eff}$ decreases. Consequently, more co-annihilation contribution is required for compensation, necessitating a decrease in $\Delta M$. The region below this always remains under-abundant.
	In the high DM mass region ($M_{\chi_2}\gsim 100$ GeV), it becomes evident that $\lambda_{\chi}$ does not necessarily have to be large. All values of $\lambda_\chi$ are permissible in this region. This is because even if $\lambda_\chi$ is small enough that DM annihilation to $\nu_R$ is not highly efficient, the relic density can still be brought to the correct range through various co-annihilation processes. 
	
This understanding gains further support through an analysis of the results as depicted in the right panel of Fig.~\ref{fig:c1relic1} by recasting the same points (as shown in the left panel of Fig~\ref{fig:c1relic1}), in the plane of $\Delta M_1$ vs $M_{\chi_2}$. By establishing a mass hierarchy among the particles in the dark sector as $M_{\chi_2} < M_{\phi_2} < M_{\phi_1}< M_{\chi_1}$, the co-annihilation processes involving $\phi$ exert a substantial influence in achieving the correct relic density. Similar to the observations in the left panel of Fig.~\ref{fig:c1relic1}, as the mass splitting $\Delta M_1 (=M_{\phi_2}-M_{\chi_2})$ increases, $\lambda_\chi$ must be augmented to offset the decrease in $\langle \sigma v\rangle_{\rm eff}$ and thus attain the correct relic density. However, when $\Delta M_{1}$ is small ($\Delta M_{1}<15$ GeV), $\lambda_{\chi}$ loses its significance, as the relic density is predominantly determined by the co-annihilation processes involving $\phi_{1,2}$. This also clarifies the observed under-abundance in the region below $\Delta M\lsim 15$ GeV in the left panel of Fig.~\ref{fig:c1relic1}, since in this scenario, co-annihilation processes involving $\phi_{1,2}$ become overwhelmingly dominant (as $\Delta M_1 < \Delta M$), leading to a significant increase in $\langle \sigma v \rangle_{\rm eff}$.

	\vspace{0.3cm}
	\noindent\textbf{Case-2 [ $10^{-7}<\lambda_\chi \leq 10^{-3} $]:}
	
\begin{figure}[h]
		\centering
		\includegraphics[scale=0.4]{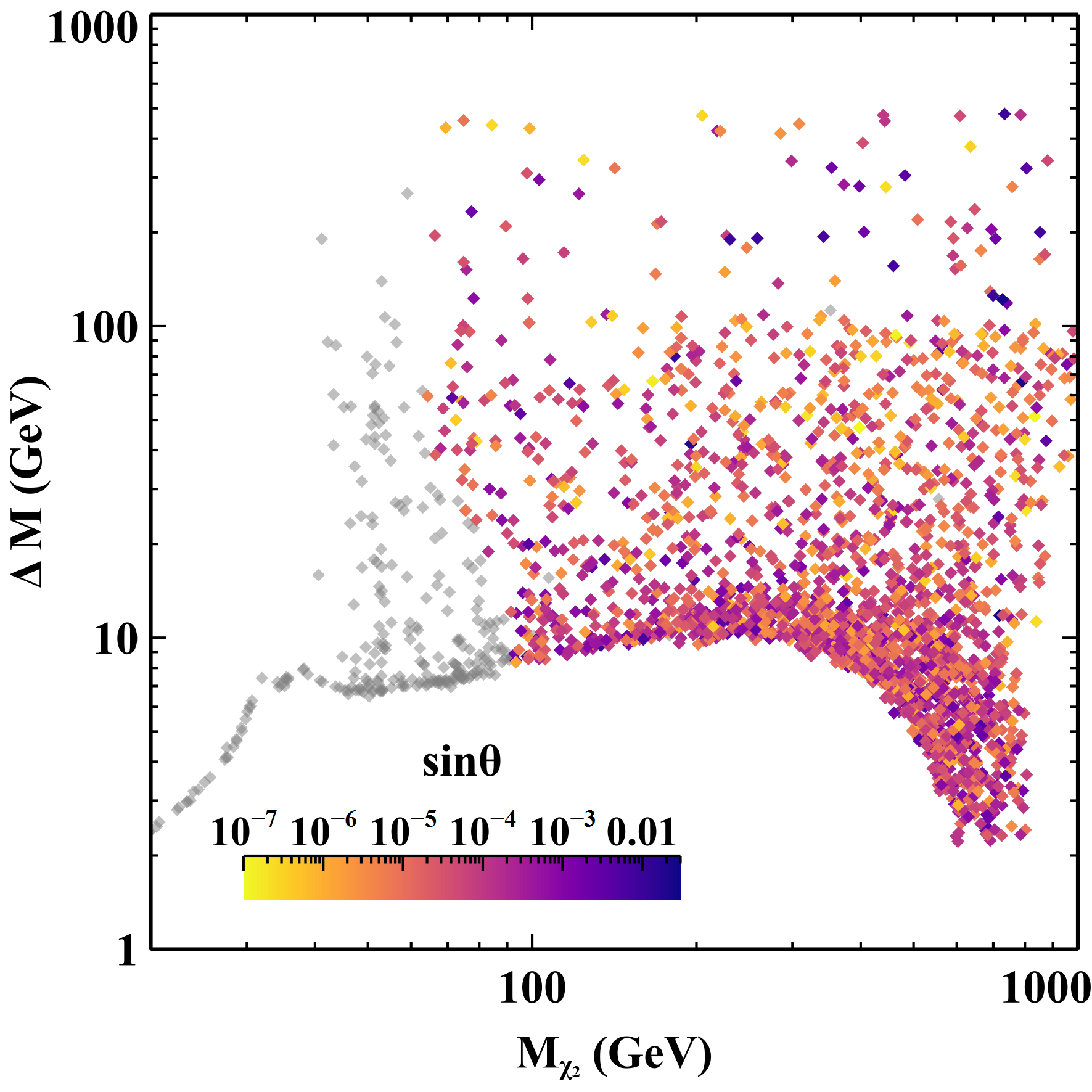}
		\includegraphics[scale=0.4]{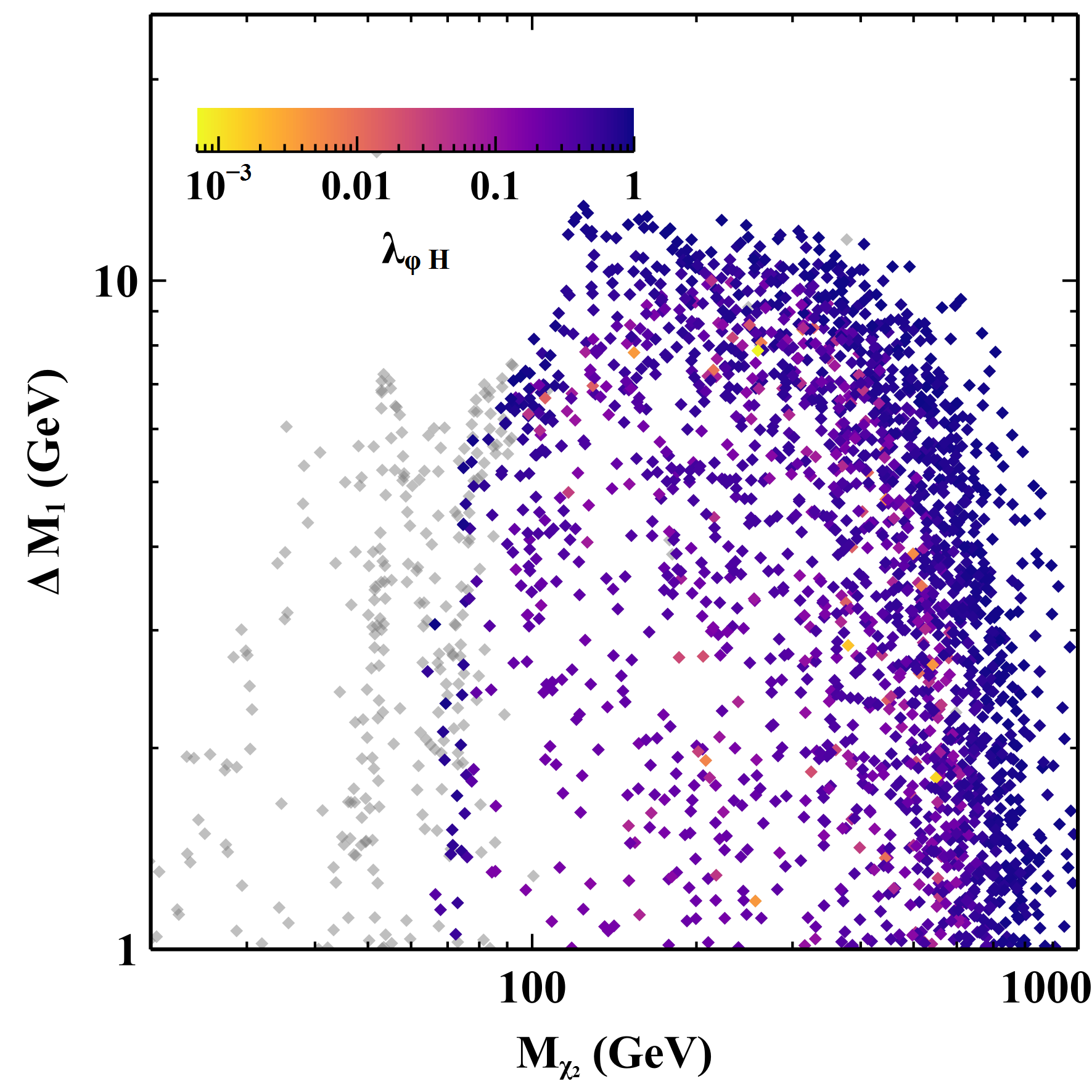}
		\caption{Points satisfying correct relic density and direct detection constraints are shown in the plane of $M_{\chi_2}$ and $\Delta M$([Left])  and that in the plane of $M_{\chi_2}$ and $\Delta M_1$([Right]). The colour code depicts the value of  $\sin\theta$ and $\lambda_{\phi H}$ in the left and right panel figures, respectively.}
		\label{fig:c2relic1}
	\end{figure}

In this case, $\lambda_\chi$ being very small ($\lambda_\chi<10^{-3}$), it does not affect the relic density of DM. Rather, DM relic density is completely decided by the co-annihilation of DM with other dark sector particles. As discussed in section~\ref{subsec:dd}, the stringent limit on $\sin\theta$ is $\sin\theta \lsim 0.03$ to be consistent with the direct detection constraints.  In the most minimal setup of fermionic singlet-doublet DM~\cite{Bhattacharya:2018fus, Borah:2021khc}, this crucially crunches the allowed parameter space to $\Delta M <20$ GeV or so to satisfy correct relic density while being consistent with the direct detection constraints. However, in our setup, we get an enhanced parameter space as compared to this because of the additional channels involving $\phi_{1,2}$ that assist in achieving the correct relic density. Moreover, these points are also not restricted by the direct detection constraints. We show the the co-annihilation effects of $\chi_1$, $\psi^-$ and $\phi_{1,2}$ in the plane of $\Delta M$ vs $M_{\chi_2}$ in the left panel of Fig.\ref{fig:c2relic1}. The same points are recasted in the plane of $\Delta M_1$ vs $M_{\chi_2}$ in the right panel of Fig.\ref{fig:c2relic1}. When $\sin\theta$ is small and $\Delta M$ is large, the singlet-doublet co-annihilation is not very effective but $\phi_{1,2}$ co-annihilation dominates over the singlet-doublet co-annihilation depending on $\lambda_{\phi H}$ coupling. It is evident from the right panel of Fig.~\ref{fig:c2relic1}, that when $\lambda_{H\Phi}$ is $\mathcal{O}(1)$, co-annihilations processes involving $\phi_{1,2}$ can significantly contribute to the correct relic density of DM. In either case, the grey coloured points are ruled out due to the LEP constraint.
	
\section{Contribution to $\Delta N_{\rm eff}$}
\label{section:delneff}
We now focus on the phenomenological consequences that can occur due to the presence of the right-handed counterparts of the SM neutrinos. As mentioned in the earlier section, the Dirac nature of neutrinos necessitates the newly added right-chiral fermions to be as light as the left-handed SM neutrinos. The presence of such additional ultra-light species in the early universe can contribute substantially to the total radiation energy density and hence to the effective relativistic degrees of freedom, $N_{\rm eff}$ which is usually parameterized as 
	\begin{equation}
	N_{\rm eff} \equiv \frac{\rho_{\rm rad}-\rho_{\gamma}}{\rho_{\nu_L}}, 
	\label{eq:Neff}
	\end{equation}
	where $\rho_{\rm rad}$ is the total energy density of the thermal plasma whereas $\rho_{\gamma}$ and $\rho_{\nu_L}$ are the energy density of photon and single active neutrino species respectively. Without the presence of any new light degrees of freedom, $N_{\rm eff}$ in SM have been calculated very precisely and quoted as $3.045$\footnote{The deviation from 3 is due to various effects like non-instantaneous neutrino decoupling, flavour
		oscillations, and finite temperature QED corrections to the electromagnetic plasma \cite{Mangano:2005cc,Grohs:2015tfy,deSalas:2016ztq,Cielo:2023bqp, Akita:2020szl, Froustey:2020mcq, Bennett:2020zkv}. } \cite{Mangano:2005cc,Grohs:2015tfy,deSalas:2016ztq,Cielo:2023bqp, Akita:2020szl, Froustey:2020mcq, Bennett:2020zkv}. The current data from the observation of cosmic microwave background (CMB) by the Planck satellite \cite{Planck:2018vyg} suggests $N_{\rm eff}=2.99^{0.34}_{-0.33}$ at $95\%$ CL (including the baryon acoustic oscillation (BAO) data) which agrees with the SM prediction. The upcoming experiments such as CMB-S4 \cite{Abazajian:2019eic}, SPT-3G \cite{SPT-3G:2019sok} are expected to be extremely sensitive to $N_{\rm eff}$ and put much more stringent bounds than the Planck experiment due to their potential of exploring all the way down to $\Delta{N_{\rm eff}} = N_{\rm eff}-N_{\rm eff}^{\rm SM}=0.06$.
	
\begin{figure}[h]
		\centering
		\includegraphics[scale=0.09]{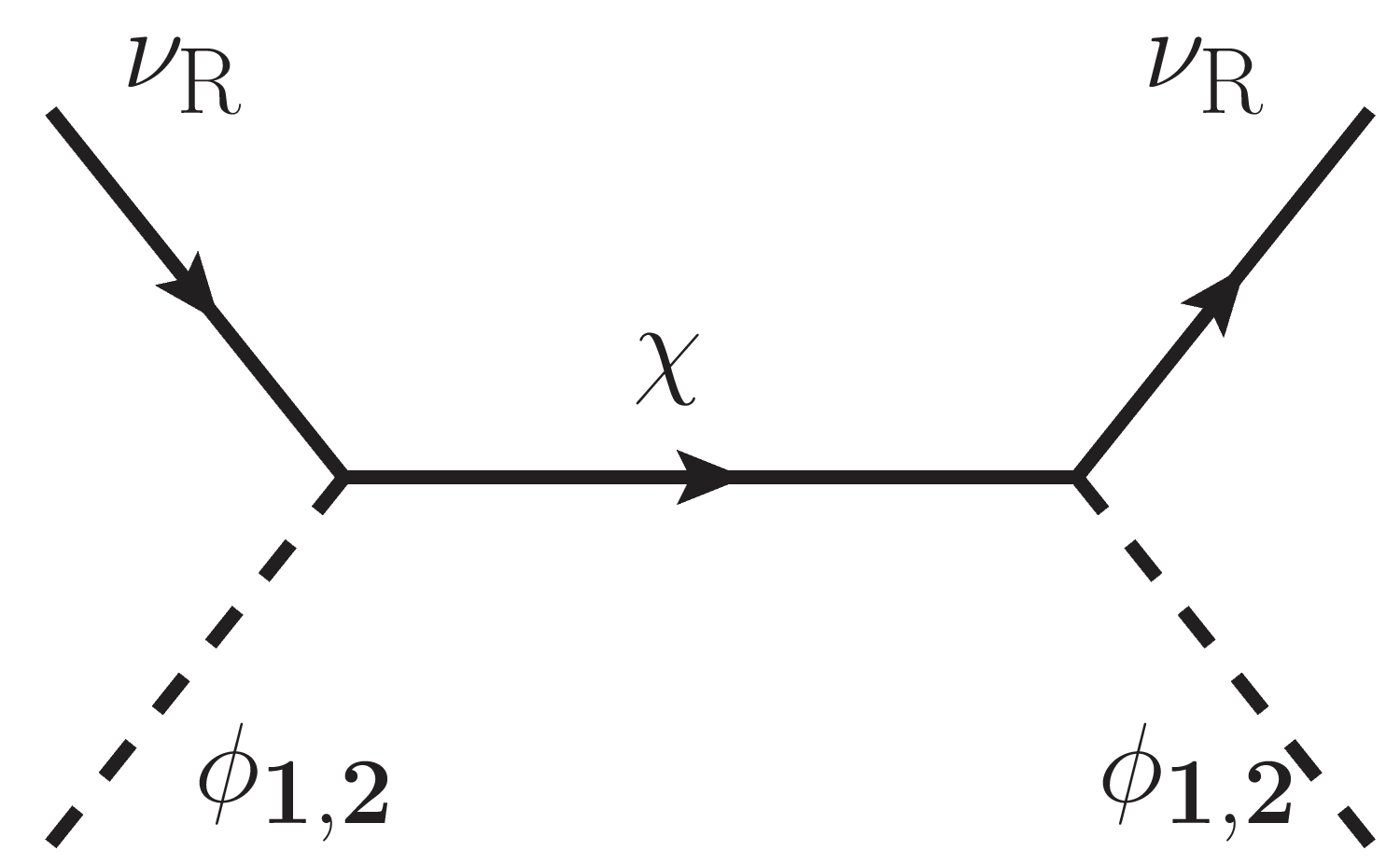}
		\includegraphics[scale=0.09]{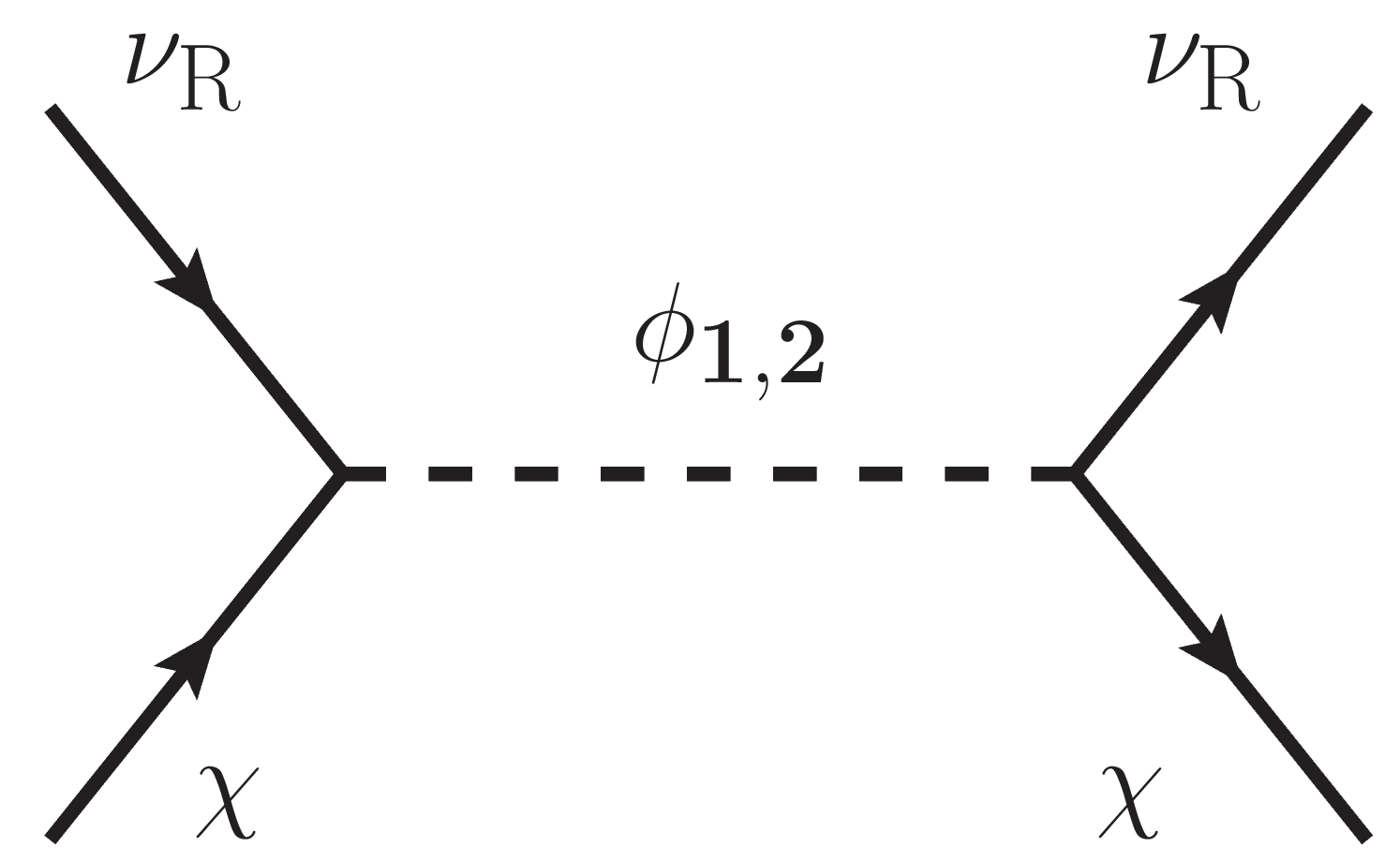}
		\caption{Feynman diagram to keep $\nu_R$ in thermal equilibrium.}
		\label{fig:image4}
	\end{figure}

	Therefore, $N_{\rm eff}$ is a quantity of immense importance to test the presence of physics beyond SM which can affect CMB and is going to play a crucial role in our discussion. The significance of such contribution may vary depending on whether such light $\nu_R$ was present in the thermal bath or produced non-thermally \cite{Luo:2020sho, Luo:2020fdt, Biswas:2022fga}. The connection between the Dirac nature of neutrinos and the origin of DM production has been studied before in \cite{Biswas:2021kio,Biswas:2022vkq} in the context of $\Delta{N_{\rm eff}}$. In this model, $\nu_R$ interacts only via Yukawa coupling $\lambda_{\chi} \overline{\nu_{R}} \phi \chi$ as shown in the Eq. \eqref{eq:lag}. Hence, the production of $\nu_R$, whether they were present in the thermal plasma or produced non-thermally with the evolution of our Universe, solely depends on $\lambda_{\chi}$. Since we have assumed $\lambda_\chi$ values for the second generation $Z_2$ odd fermions is much smaller than that of the first generation, the contribution of second generation fermions to $\Delta N_{\rm eff}$ is substantially smaller than the first generation. Therefore, we study the contribution from the first generation $Z_2$ fermions to $\Delta{N_{\rm eff}}$ as mentioned by case-1 and case-2 in section \ref{section:DMpheno}. From Eq. \eqref{eq:Neff}, the additional contribution to $N_{\rm eff}$ at the time of CMB coming from the presence of $\nu_R$ in the total radiation energy density can be written as,
	\begin{eqnarray}
	\Delta{N_{\rm eff}}&=& N_{\nu_R}\times \frac{\rho_{\nu_R}}{\rho_{\nu_L}} \Bigg|_{\rm T=T_{\rm CMB}},
	\end{eqnarray}
	where $N_{\nu_R}$ is the number of relativistic $\nu_R $ species, and $\rho_{\nu_R}$ is the energy density of the single $\nu_R$.  In the above
	equation, it is assumed that all three $\nu_R$ behave identically, hence contribute equally to the energy density. Depending on the production mechanism of $\nu_R$ we will discuss two following cases:

	{\bf Case-1[$\lambda_{\chi}\, >\, 10^{-3}$]:}\\
	In this case, the $\nu_R$ can be thermally produced in the early universe from the annihilation of $\chi$ and $\phi$. Fig. \ref{fig:image4} shows the interactions responsible for keeping $\nu_R$ in equilibrium. Once these interaction rates dropped below the expansion rate of the universe, $\nu_R$ became decoupled from the thermal plasma and their temperature evolved independently. Hence by using entropy conservation in both sectors, one can write the excess contribution as 
	\begin{equation}\label{eq:7c}
	\Delta N_{\rm eff}=N_{\nu_R}\times\left(\frac{T_{\nu_R}}{T_{\nu_L}}\right)^4=N_{\nu_R}\left(\frac{g_{*s}(T^{\rm dec}_{\nu_L})}{g_{*s}(T^{\rm dec}_{\nu_R})}\right)^{4/3},
	\end{equation} 
	where $g_{*s}(T_{\kappa}^{\rm dec})$ is the relativistic entropy degrees of freedom at temperature $T_{\kappa}^{\rm dec}$ which is the decoupling temperature of the species $\kappa(=\nu_L, \nu_R )$ from the thermal bath. One important thing to keep in mind is that $\lambda_{\chi}$ also plays an important role in neutrino mass generation. Increasing $\lambda_{\chi}$ corresponds to a smaller mixing angle $\sin\theta$ to satisfy the correct neutrino mass. The viable parameter space for $\Delta N_{\rm eff}$ is obtained by satisfying all possible constraints such as neutrino mass, $(g-2)_\mu$ and DM relic density while satisfying direct detection constraints.

\begin{figure}[h]
		\centering
		\includegraphics[scale=0.45]{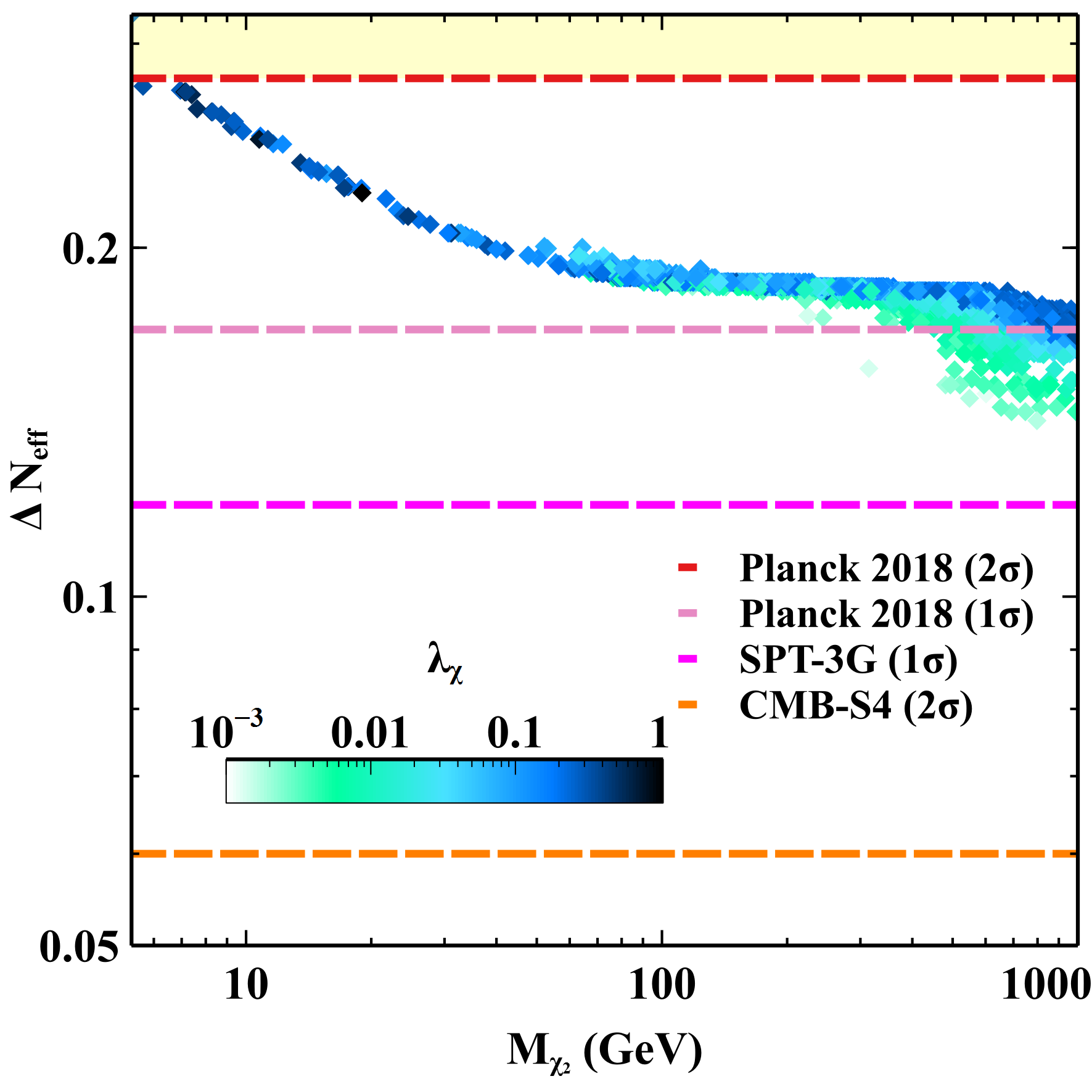}
		\caption{$\Delta N_{\rm eff}$ vs DM mass and the colour code is used for $\lambda_\chi$. The dashed lines show experimental sensitivity, and the shaded region is excluded by Planck 2018 with 2$\sigma$ sensitivity.}
		\label{fig:deln1}
	\end{figure}

	Fig. \ref{fig:deln1} shows the variation of $\Delta N_{\rm eff}$ as a function of DM mass ranging from 1 GeV to 1 TeV. The colour gradient shows the dependence on $\lambda_\chi$. One can clearly see a correlation between $\Delta{N_{\rm eff}}$, $\lambda_{\chi}$ and the mass of DM. Larger $\lambda_{\chi}$ and lighter $M_{\chi_2}$ corresponds to larger contribution to the $\Delta{N_{\rm eff}}$. This can be understood as follows. A larger $\lambda_{\chi}$ or lighter $M_{\chi_2}$ forces $\chi_2$ and hence $\nu_R$ to be in thermal bath for longer time. The late decoupling of $\nu_R$ from the plasma decreases the denominator in Eq. \eqref{eq:7c} and increases its contribution to the $N_{\rm eff}$. The yellow-shaded region is already excluded from the PLANCK 2018 data at 2$\sigma$ CL. More importantly, future observations of microwave background can test this scenario fully as shown by the pink and orange dashed line, which corresponds to SPT-3G and CMB-S4, respectively. This is because once three of the $\nu_R$s were produced in the thermal bath of the early Universe, $\Delta{N_{\rm eff}}$ would always have some minimum contribution of 0.14 which is well above the future prediction from CMB-S4 or SPT-3G. This is the manifestation of the conservation of entropy.\\
	
	{\bf Case-2[ $10^{-7}<\lambda_\chi \leq 10^{-3} $]:}
	\\
	\begin{figure}[h]
		\centering
		\includegraphics[scale=0.1]{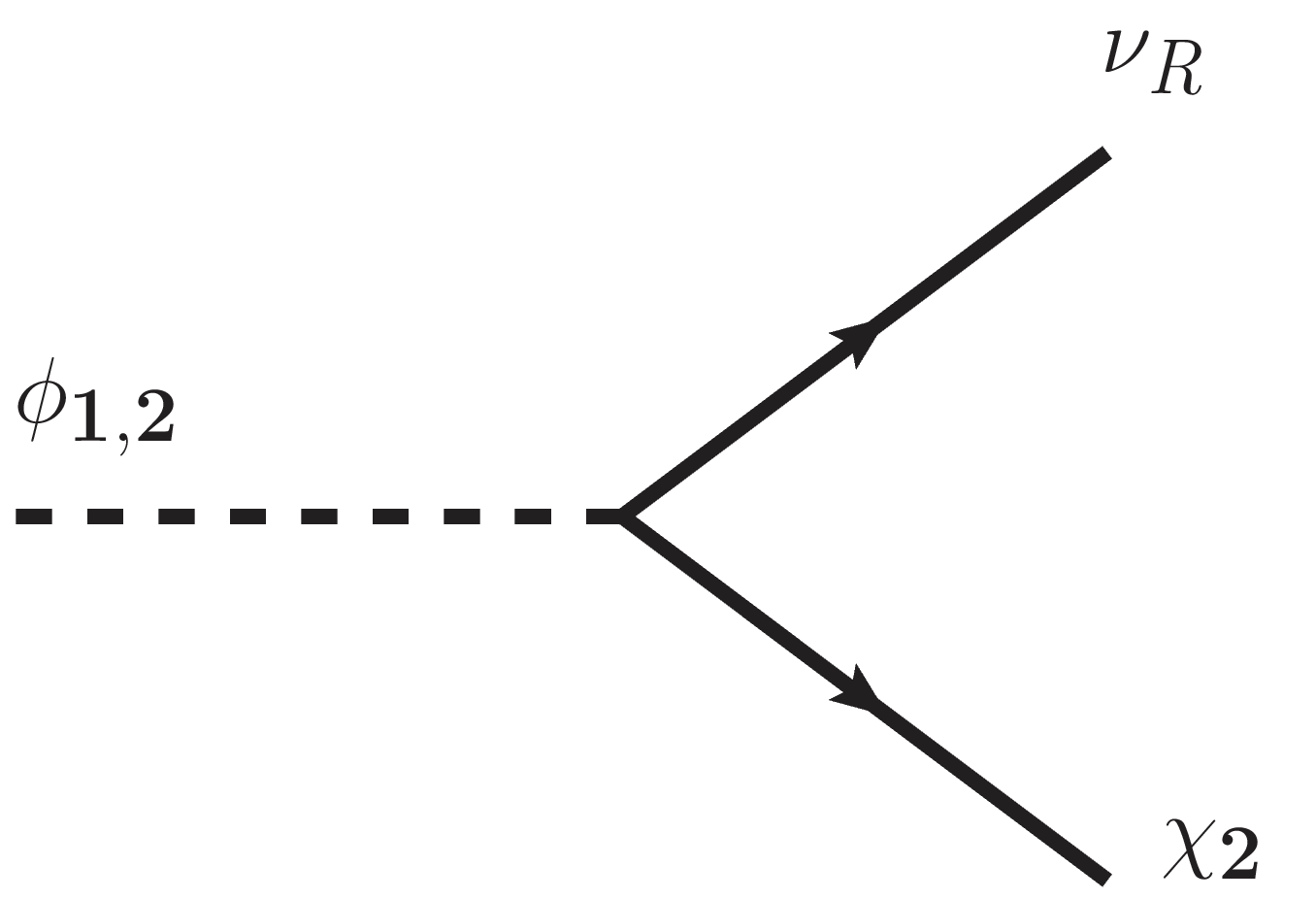}
		\caption{Production of $\nu_R$ from the decay of $\phi_{1,2}$.}
		\label{fig:image5}
	\end{figure}
	
	\begin{figure}[h]
		\centering
		\includegraphics[scale=0.45]{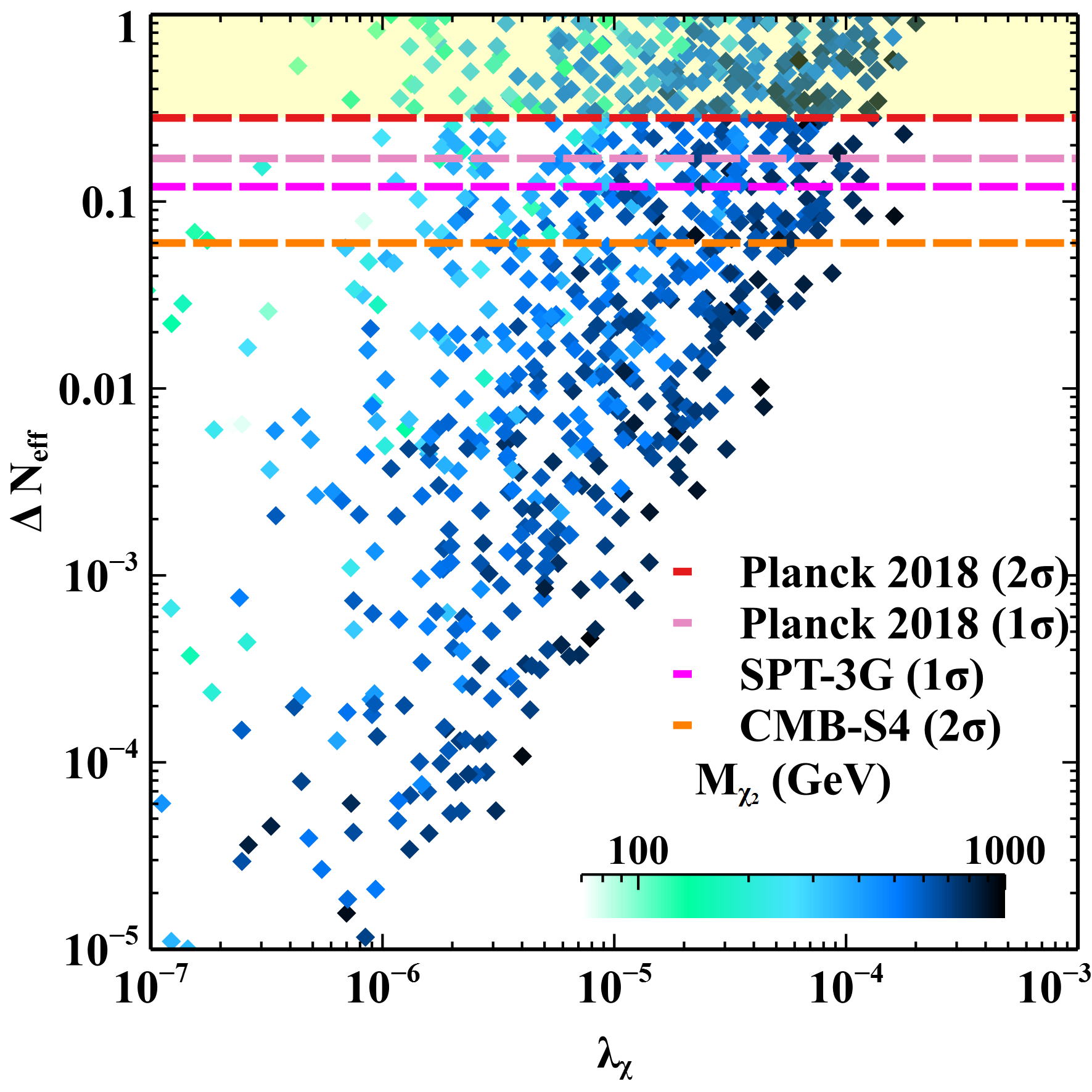}
		\caption{$\Delta N_{\rm eff}$ vs $\lambda_\chi$. The colour code is shown for DM mass. Dashed lines show the experimental sensitivity and the shaded region is for the excluded region.}
		\label{fig:deln2}
	\end{figure}
As discussed above, once the $\nu_R$s or any light degrees of freedom were thermalized with the SM plasma, there would always be some minimum contribution to the $\Delta{N_{\rm eff}}$. However, this is not the same in the case of non-thermal production. As we go to the smaller $\lambda_\chi$, $\nu_R$ would no longer be produced in the thermal bath. Rather, they would be produced from the decay of ${\phi_{1,2}}$ with a $\chi_2$ as shown in Fig. \ref{fig:image5}. In this case, we also need to track the evolution of ${\phi_{1,2}}$ as that will be important to calculate the total energy injected into the radiation energy in terms of light $\nu_R$. We assume that ${\phi_{1,2}}$ was present in the thermal bath of the early universe and decays both from equilibrium and after freezing out from the thermal bath.
	So, to track the evolution of $\nu_R$ and ${\phi_{i}}(i=1,2)$ we need to solve the following Boltzmann equations,
	\begin{equation*}
	\frac{dY_{\phi_i}}{dx}=\frac{\beta s}{Hx}\left[-\langle\sigma v\rangle_{{\phi_i} {\phi_i}\rightarrow X \bar{X}}(Y^2_{\phi_i}-(Y^{\rm eq}_{\phi_i})^2)-\frac{\Gamma_{\phi_i}}{s}\frac{K_1(x)}{K_2(x)}Y_{\phi_i}\right]
	\end{equation*}
	\begin{equation}
	\frac{d\Tilde{Y}_{\nu_R}}{dx}=\frac{\beta}{Hs^{1/3}x}\langle E\Gamma\rangle Y_{\phi_{i}},
	\label{eq:11}
	\end{equation}
	
where the dimensionless parameters $Y_{\phi_i}=\sum_{i=1}^2{n_{\phi_i}}/{s} ~{\rm and}~\Tilde{Y}_{\nu_R}={\rho_{\nu_R}}/{s^{4/3}}$. In the above equation $\Gamma_{\phi_i}$ and $\langle E\Gamma\rangle$ can be expressed as
\begin{eqnarray}
\Gamma({{\phi_i}\rightarrow\chi \nu_R})&=&\frac{1}{8\pi}\lambda_\chi^2 \cos\theta^2 M_{\phi_i} \left(1-\frac{M_{\chi_2}^2}{M_{\phi_i^2}}\right)^2\\
\langle E\Gamma({{\phi_i}\rightarrow\chi \nu_R})\rangle&=&\frac{1}{16\pi}\lambda_\chi^2 \cos\theta^2 M_{\phi_i}^2 \left(1-\frac{M_{\chi_2}^2}{M_{\phi_i^2}}\right)^3. 
\label{eq:15}
\end{eqnarray}	
	

After solving the above Boltzmann equations, $\Delta N_{\rm eff}$ produced by $\rho_{\nu_R}$ can be calculated as:
	\begin{eqnarray}\label{eq:deln2}
	\Delta N_{\rm eff}&=&3 \times 2\left(\frac{\rho_{\nu_R}}{\rho_{\nu_L}}\right)_{T_D(\nu_L)\nonumber}\\
	&=&6\left(\frac{s^{4/3}\Tilde{Y}_{\nu_R}}{\rho_{\nu_L}}\right)_{T_D(\nu_L)}
	\end{eqnarray}
The factor of 3 in Eq. \eqref{eq:deln2} corresponds to the three generations of $\nu_R$ and 2 incorporates both particles and antiparticles. Fig. \ref{fig:deln2} shows the variation of $\Delta{N_{\rm eff}}$ as a function of $\lambda_{\chi}$ where the colour bar represents the variation of $M_{\chi_2}$. We have considered the mass difference between $\phi_2$ and $\chi_2$ to be $\leq$ 10 GeV. Eq. \eqref{eq:15} says that with larger $\lambda_\chi$, the decay width of $\phi_{1,2}$ will increase and hence more RHNs will be produced. If we increase the DM mass, the factor inside the bracket will decrease, hence the decay width will decrease. We can see $\Delta N_{\rm eff}$ increases with $\lambda_\chi$ and for a fixed $\lambda_\chi$, it decreases with increase in $M_{\chi_2}$. One can note that some part of the parameter space is excluded from the present PLANCK data and future experiments can probe some part of it. However, for smaller $\lambda_{\chi}$, $\Delta{N_{\rm eff}}$ becomes extremely small and drops well below the future projections of CMB-S4 or SPT-3G limits.  
	
	
\section{Conclusion}
\label{section:conclusion}	
	
In this study, we delved into a singlet-doublet fermionic DM model that offers a scotogenic avenue for generating Dirac neutrino masses. To achieve this, we extended the basic singlet-doublet DM framework by introducing an additional singlet scalar and the right-chiral components of neutrinos, $\nu_R$s. This extension also aids in addressing the $(g-2)_\mu$ anomaly, with a positive contribution arising from the one-loop diagram mediated by the charged fermion doublet $\psi^{-}$ and $\phi_{1,2}$. Following the inclusion of constraints stemming from neutrino mass, $(g-2)_\mu$, and lepton flavour violation (LFV), we conducted a comprehensive exploration of the DM phenomenology.
	
In the analysis of DM, we delineated two distinct cases for examination, based on the thermalisation criteria of $\nu_R$, as this can yield intriguing implications from the perspective of the additional effective neutrino species ($\Delta N_{\rm eff}$). We find an enhanced parameter space because of the presence of $\phi_{1,2}$ in the dark sector as it helps in achieving correct relic density due to additional co-annihilation contributions. We considered both tree-level and loop-level DM-nucleon scattering possibilities for the DM direct detection perspectives. While direct DM search experiments do not impose stringent restrictions on the model parameters, $\Delta N_{\rm eff}$ offers an additional cosmological probe for the model. As $\nu_R$ is connected to the thermal bath, only through the dark sector particles, in the scenario, where $\nu_R$ is thermalized, future CMB experiment SPT-3G can probe the higher DM mass range of the model whereas when $\nu_R$ is produced non-thermally, both SPT-3G and CMB-S4 can probe significant portion of the model parameter space.

\section*{Acknowledgements}
The work of DB is supported by the Science and Engineering Research Board (SERB), Government of India grant MTR/2022/000575. SM acknowledges the financial support from the National Research Foundation of Korea grant 2022R1A2C1005050. The work of DN is supported by the National Research Foundation of Korea (NRF) grants, grant no. 2019R1A2C3005009(DN). DN also thanks the Department of Theoretical Physics, CERN where he was a visitor during the completion of the work.  The work of NS is supported by the Department of Atomic Energy-Board of Research in Nuclear 
	Sciences, Government of India (Ref. Number: 58/14/15/2021- BRNS/37220). SS would like to thank Pritam Das for useful discussion.

\appendix	
\section{Lagrangian}
	Following from Eq. \eqref{eq:lag}, we can write the Lagrangian in terms of physical states as, 
	
	\begin{eqnarray}
	\mathcal{L}^{\rm DM}_{\rm int} &=& i \overline{\psi} \gamma^\mu (\partial_\mu-g\frac{\tau_i}{2} W^i_\mu-g^{'}\frac{Y}{2} B_\mu) \Psi + i \overline{\chi} \gamma^\mu \partial_\mu \chi- (\lambda_{\psi}\overline{L} \phi \Psi+\lambda_{\chi} \overline{\nu_{R}} \phi  \chi +{\rm h.c.})\nonumber\\
	\mathcal{L}^{\rm DM}_{\rm int} &=& g_Z \Big[\sin^2\theta \overline{\chi_2}\gamma^{\mu}Z_{\mu}\chi_2+\cos^2\theta \overline{\chi_1}\gamma^{\mu}Z_{\mu}\chi_1 
	+\sin\theta \cos\theta(\overline{\chi_1}\gamma^{\mu}Z_{\mu}\chi_2+\overline{\chi_2}\gamma^{\mu}Z_{\mu}\chi_1)\Big]   \nonumber \\
	&+&g_W(\sin\theta \overline{\chi_2}\gamma^\mu W_\mu^+ \psi^- + \cos\theta \overline{\chi_1}\gamma^\mu W_\mu^+ \psi^- )  +g_W( \sin\theta{\psi^+}\gamma^\mu W_\mu^- \chi_2 + \cos\theta {\psi^+}\gamma^\mu W_\mu^- \chi_1 ) \nonumber \\
	&-& g_Z ~c_{2w}~ {\psi^+}\gamma^{\mu}Z_{\mu}\psi^- - e_0 {\psi^+}\gamma^{\mu}A_{\mu}\psi^- -\lambda_\psi \overline{e_L} \phi \psi^-  \nonumber\\
	&-&\Big[ \cos\theta(\lambda_\psi \overline{\nu_L}\phi\chi_1+\lambda_\chi\overline{\nu_R}\phi\chi_2)+\sin\theta(\lambda_\chi\overline{\nu_R}\phi\chi_1-\lambda_\psi \overline{\nu_L}\phi\chi_2)\Big]
	\end{eqnarray}
	where $g_Z=e_0/2 s_w c_w$, $g_W=e_0/\sqrt{2} s_w$ and $s_w$, $c_w, c_{2w}$ denote $\sin \theta_W$ and $\cos \theta_W$ and $\cos2\theta_{W}$ respectively.
	


\begin{thebibliography}{100}
	
	\bibitem{Planck:2018vyg}
	{\scshape Planck} collaboration, \emph{{Planck 2018 results. VI. Cosmological
			parameters}},
	\href{https://doi.org/10.1051/0004-6361/201833910}{\emph{Astron. Astrophys.}
		{\bfseries 641} (2020) A6}
	[\href{https://arxiv.org/abs/1807.06209}{{\ttfamily 1807.06209}}].
	
	\bibitem{WMAP:2012nax}
	{\scshape WMAP} collaboration, \emph{{Nine-Year Wilkinson Microwave Anisotropy
			Probe (WMAP) Observations: Cosmological Parameter Results}},
	\href{https://doi.org/10.1088/0067-0049/208/2/19}{\emph{Astrophys. J. Suppl.}
		{\bfseries 208} (2013) 19} [\href{https://arxiv.org/abs/1212.5226}{{\ttfamily
			1212.5226}}].
	
	\bibitem{Lee:1977ua}
	B.W.~Lee and S.~Weinberg, \emph{{Cosmological Lower Bound on Heavy Neutrino
			Masses}}, \href{https://doi.org/10.1103/PhysRevLett.39.165}{\emph{Phys. Rev.
			Lett.} {\bfseries 39} (1977) 165}.
	
	\bibitem{Griest:1989wd}
	K.~Griest and M.~Kamionkowski, \emph{{Unitarity Limits on the Mass and Radius
			of Dark Matter Particles}},
	\href{https://doi.org/10.1103/PhysRevLett.64.615}{\emph{Phys. Rev. Lett.}
		{\bfseries 64} (1990) 615}.
	
	\bibitem{Super-Kamiokande:1998kpq}
	{\scshape Super-Kamiokande} collaboration, \emph{{Evidence for oscillation of
			atmospheric neutrinos}},
	\href{https://doi.org/10.1103/PhysRevLett.81.1562}{\emph{Phys. Rev. Lett.}
		{\bfseries 81} (1998) 1562}
	[\href{https://arxiv.org/abs/hep-ex/9807003}{{\ttfamily hep-ex/9807003}}].
	
	\bibitem{SNO:2001kpb}
	{\scshape SNO} collaboration, \emph{{Measurement of the rate of $\nu_e+d \to
			p+p+e^-$ interactions produced by $^8$B solar neutrinos at the Sudbury
			Neutrino Observatory}},
	\href{https://doi.org/10.1103/PhysRevLett.87.071301}{\emph{Phys. Rev. Lett.}
		{\bfseries 87} (2001) 071301}
	[\href{https://arxiv.org/abs/nucl-ex/0106015}{{\ttfamily nucl-ex/0106015}}].
	
	\bibitem{DoubleChooz:2011ymz}
	{\scshape Double Chooz} collaboration, \emph{{Indication of Reactor
			$\bar{\nu}_e$ Disappearance in the Double Chooz Experiment}},
	\href{https://doi.org/10.1103/PhysRevLett.108.131801}{\emph{Phys. Rev. Lett.}
		{\bfseries 108} (2012) 131801}
	[\href{https://arxiv.org/abs/1112.6353}{{\ttfamily 1112.6353}}].
	
	\bibitem{DayaBay:2012fng}
	{\scshape Daya Bay} collaboration, \emph{{Observation of electron-antineutrino
			disappearance at Daya Bay}},
	\href{https://doi.org/10.1103/PhysRevLett.108.171803}{\emph{Phys. Rev. Lett.}
		{\bfseries 108} (2012) 171803}
	[\href{https://arxiv.org/abs/1203.1669}{{\ttfamily 1203.1669}}].
	
	\bibitem{RENO:2012mkc}
	{\scshape RENO} collaboration, \emph{{Observation of Reactor Electron
			Antineutrino Disappearance in the RENO Experiment}},
	\href{https://doi.org/10.1103/PhysRevLett.108.191802}{\emph{Phys. Rev. Lett.}
		{\bfseries 108} (2012) 191802}
	[\href{https://arxiv.org/abs/1204.0626}{{\ttfamily 1204.0626}}].
	
	\bibitem{ParticleDataGroup:2022pth}
	{\scshape Particle Data Group} collaboration, \emph{{Review of Particle
			Physics}}, \href{https://doi.org/10.1093/ptep/ptac097}{\emph{PTEP} {\bfseries
			2022} (2022) 083C01}.
	
	\bibitem{Biswas:2021kio}
	A.~Biswas, D.~Borah and D.~Nanda, \emph{{Light Dirac neutrino portal dark
			matter with observable \ensuremath{\Delta}Neff}},
	\href{https://doi.org/10.1088/1475-7516/2021/10/002}{\emph{JCAP} {\bfseries
			10} (2021) 002} [\href{https://arxiv.org/abs/2103.05648}{{\ttfamily
			2103.05648}}].
	
	\bibitem{Borah:2022obi}
	D.~Borah, S.~Mahapatra, D.~Nanda and N.~Sahu, \emph{{Type II Dirac seesaw with
			observable \ensuremath{\Delta}Neff in the light of W-mass anomaly}},
	\href{https://doi.org/10.1016/j.physletb.2022.137297}{\emph{Phys. Lett. B}
		{\bfseries 833} (2022) 137297}
	[\href{https://arxiv.org/abs/2204.08266}{{\ttfamily 2204.08266}}].
	
	\bibitem{Biswas:2022vkq}
	A.~Biswas, D.~Borah, N.~Das and D.~Nanda, \emph{{Freeze-in dark matter via a
			light Dirac neutrino portal}},
	\href{https://doi.org/10.1103/PhysRevD.107.015015}{\emph{Phys. Rev. D}
		{\bfseries 107} (2023) 015015}
	[\href{https://arxiv.org/abs/2205.01144}{{\ttfamily 2205.01144}}].
	
	\bibitem{Ma:2016mwh}
	E.~Ma and O.~Popov, \emph{{Pathways to Naturally Small Dirac Neutrino Masses}},
	\href{https://doi.org/10.1016/j.physletb.2016.11.027}{\emph{Phys. Lett. B}
		{\bfseries 764} (2017) 142}
	[\href{https://arxiv.org/abs/1609.02538}{{\ttfamily 1609.02538}}].
	
	\bibitem{Yao:2017vtm}
	C.-Y.~Yao and G.-J.~Ding, \emph{{Systematic Study of One-Loop Dirac Neutrino
			Masses and Viable Dark Matter Candidates}},
	\href{https://doi.org/10.1103/PhysRevD.96.095004}{\emph{Phys. Rev. D}
		{\bfseries 96} (2017) 095004}
	[\href{https://arxiv.org/abs/1707.09786}{{\ttfamily 1707.09786}}].
	
	\bibitem{Carvajal:2018ohk}
	C.D.R.~Carvajal and O.~Zapata, \emph{{One-loop Dirac neutrino mass and mixed
			axion-WIMP dark matter}},
	\href{https://doi.org/10.1103/PhysRevD.99.075009}{\emph{Phys. Rev. D}
		{\bfseries 99} (2019) 075009}
	[\href{https://arxiv.org/abs/1812.06364}{{\ttfamily 1812.06364}}].
	
	\bibitem{Jana:2019mgj}
	S.~Jana, P.K.~Vishnu and S.~Saad, \emph{{Minimal realizations of Dirac neutrino
			mass from generic one-loop and two-loop topologies at $d = 5$}},
	\href{https://doi.org/10.1088/1475-7516/2020/04/018}{\emph{JCAP} {\bfseries
			04} (2020) 018} [\href{https://arxiv.org/abs/1910.09537}{{\ttfamily
			1910.09537}}].
	
	\bibitem{Nanda:2019nqy}
	D.~Nanda and D.~Borah, \emph{{Connecting Light Dirac Neutrinos to a
			Multi-component Dark Matter Scenario in Gauged $B-L$ Model}},
	\href{https://doi.org/10.1140/epjc/s10052-020-8122-4}{\emph{Eur. Phys. J. C}
		{\bfseries 80} (2020) 557}
	[\href{https://arxiv.org/abs/1911.04703}{{\ttfamily 1911.04703}}].
	
	\bibitem{Borah:2020jzi}
	D.~Borah, S.~Mahapatra, D.~Nanda and N.~Sahu, \emph{{Inelastic fermion dark
			matter origin of XENON1T excess with muon $(g-2)$ and light neutrino
			mass}}, \href{https://doi.org/10.1016/j.physletb.2020.135933}{\emph{Phys.
			Lett. B} {\bfseries 811} (2020) 135933}
	[\href{https://arxiv.org/abs/2007.10754}{{\ttfamily 2007.10754}}].
	
	\bibitem{Biswas:2022fga}
	A.~Biswas, D.K.~Ghosh and D.~Nanda, \emph{{Concealing Dirac neutrinos from
			cosmic microwave background}},
	\href{https://doi.org/10.1088/1475-7516/2022/10/006}{\emph{JCAP} {\bfseries
			10} (2022) 006} [\href{https://arxiv.org/abs/2206.13710}{{\ttfamily
			2206.13710}}].
	
	\bibitem{Wang:2016lve}
	W.~Wang and Z.-L.~Han, \emph{{Naturally Small Dirac Neutrino Mass with
			Intermediate $SU(2)_{L}$ Multiplet Fields}},
	\href{https://doi.org/10.1007/JHEP04(2017)166}{\emph{JHEP} {\bfseries 04}
		(2017) 166} [\href{https://arxiv.org/abs/1611.03240}{{\ttfamily
			1611.03240}}].
	
	\bibitem{Aoyama:2012wk}
	T.~Aoyama, M.~Hayakawa, T.~Kinoshita and M.~Nio, \emph{{Complete Tenth-Order
			QED Contribution to the Muon g-2}},
	\href{https://doi.org/10.1103/PhysRevLett.109.111808}{\emph{Phys. Rev. Lett.}
		{\bfseries 109} (2012) 111808}
	[\href{https://arxiv.org/abs/1205.5370}{{\ttfamily 1205.5370}}].
	
	\bibitem{Aoyama:2019ryr}
	T.~Aoyama, T.~Kinoshita and M.~Nio, \emph{{Theory of the Anomalous Magnetic
			Moment of the Electron}},
	\href{https://doi.org/10.3390/atoms7010028}{\emph{Atoms} {\bfseries 7} (2019)
		28}.
	
	\bibitem{Czarnecki:2002nt}
	A.~Czarnecki, W.J.~Marciano and A.~Vainshtein, \emph{{Refinements in
			electroweak contributions to the muon anomalous magnetic moment}},
	\href{https://doi.org/10.1103/PhysRevD.67.073006}{\emph{Phys. Rev. D}
		{\bfseries 67} (2003) 073006}
	[\href{https://arxiv.org/abs/hep-ph/0212229}{{\ttfamily hep-ph/0212229}}].
	
	\bibitem{Gnendiger:2013pva}
	C.~Gnendiger, D.~St\"ockinger and H.~St\"ockinger-Kim, \emph{{The electroweak
			contributions to $(g-2)_\mu$ after the Higgs boson mass measurement}},
	\href{https://doi.org/10.1103/PhysRevD.88.053005}{\emph{Phys. Rev. D}
		{\bfseries 88} (2013) 053005}
	[\href{https://arxiv.org/abs/1306.5546}{{\ttfamily 1306.5546}}].
	
	\bibitem{Davier:2017zfy}
	M.~Davier, A.~Hoecker, B.~Malaescu and Z.~Zhang, \emph{{Reevaluation of the
			hadronic vacuum polarisation contributions to the Standard Model predictions
			of the muon $g-2$ and ${\alpha (m_Z^2)}$ using newest hadronic cross-section
			data}}, \href{https://doi.org/10.1140/epjc/s10052-017-5161-6}{\emph{Eur.
			Phys. J. C} {\bfseries 77} (2017) 827}
	[\href{https://arxiv.org/abs/1706.09436}{{\ttfamily 1706.09436}}].
	
	\bibitem{Keshavarzi:2018mgv}
	A.~Keshavarzi, D.~Nomura and T.~Teubner, \emph{{Muon $g-2$ and $\alpha(M_Z^2)$:
			a new data-based analysis}},
	\href{https://doi.org/10.1103/PhysRevD.97.114025}{\emph{Phys. Rev. D}
		{\bfseries 97} (2018) 114025}
	[\href{https://arxiv.org/abs/1802.02995}{{\ttfamily 1802.02995}}].
	
	\bibitem{Colangelo:2018mtw}
	G.~Colangelo, M.~Hoferichter and P.~Stoffer, \emph{{Two-pion contribution to
			hadronic vacuum polarization}},
	\href{https://doi.org/10.1007/JHEP02(2019)006}{\emph{JHEP} {\bfseries 02}
		(2019) 006} [\href{https://arxiv.org/abs/1810.00007}{{\ttfamily
			1810.00007}}].
	
	\bibitem{Hoferichter:2019mqg}
	M.~Hoferichter, B.-L.~Hoid and B.~Kubis, \emph{{Three-pion contribution to
			hadronic vacuum polarization}},
	\href{https://doi.org/10.1007/JHEP08(2019)137}{\emph{JHEP} {\bfseries 08}
		(2019) 137} [\href{https://arxiv.org/abs/1907.01556}{{\ttfamily
			1907.01556}}].
	
	\bibitem{Davier:2019can}
	M.~Davier, A.~Hoecker, B.~Malaescu and Z.~Zhang, \emph{{A new evaluation of the
			hadronic vacuum polarisation contributions to the muon anomalous magnetic
			moment and to $\mathbf{\boldsymbol\alpha(m_Z^2)}$}},
	\href{https://doi.org/10.1140/epjc/s10052-020-7792-2}{\emph{Eur. Phys. J. C}
		{\bfseries 80} (2020) 241}
	[\href{https://arxiv.org/abs/1908.00921}{{\ttfamily 1908.00921}}].
	
	\bibitem{Keshavarzi:2019abf}
	A.~Keshavarzi, D.~Nomura and T.~Teubner, \emph{{$g-2$ of charged leptons,
			$\alpha (M^2_Z)$ , and the hyperfine splitting of muonium}},
	\href{https://doi.org/10.1103/PhysRevD.101.014029}{\emph{Phys. Rev. D}
		{\bfseries 101} (2020) 014029}
	[\href{https://arxiv.org/abs/1911.00367}{{\ttfamily 1911.00367}}].
	
	\bibitem{Kurz:2014wya}
	A.~Kurz, T.~Liu, P.~Marquard and M.~Steinhauser, \emph{{Hadronic contribution
			to the muon anomalous magnetic moment to next-to-next-to-leading order}},
	\href{https://doi.org/10.1016/j.physletb.2014.05.043}{\emph{Phys. Lett. B}
		{\bfseries 734} (2014) 144}
	[\href{https://arxiv.org/abs/1403.6400}{{\ttfamily 1403.6400}}].
	
	\bibitem{Melnikov:2003xd}
	K.~Melnikov and A.~Vainshtein, \emph{{Hadronic light-by-light scattering
			contribution to the muon anomalous magnetic moment revisited}},
	\href{https://doi.org/10.1103/PhysRevD.70.113006}{\emph{Phys. Rev. D}
		{\bfseries 70} (2004) 113006}
	[\href{https://arxiv.org/abs/hep-ph/0312226}{{\ttfamily hep-ph/0312226}}].
	
	\bibitem{Masjuan:2017tvw}
	P.~Masjuan and P.~Sanchez-Puertas, \emph{{Pseudoscalar-pole contribution to the
			$(g_{\mu}-2)$: a rational approach}},
	\href{https://doi.org/10.1103/PhysRevD.95.054026}{\emph{Phys. Rev. D}
		{\bfseries 95} (2017) 054026}
	[\href{https://arxiv.org/abs/1701.05829}{{\ttfamily 1701.05829}}].
	
	\bibitem{Colangelo:2017fiz}
	G.~Colangelo, M.~Hoferichter, M.~Procura and P.~Stoffer, \emph{{Dispersion
			relation for hadronic light-by-light scattering: two-pion contributions}},
	\href{https://doi.org/10.1007/JHEP04(2017)161}{\emph{JHEP} {\bfseries 04}
		(2017) 161} [\href{https://arxiv.org/abs/1702.07347}{{\ttfamily
			1702.07347}}].
	
	\bibitem{Hoferichter:2018kwz}
	M.~Hoferichter, B.-L.~Hoid, B.~Kubis, S.~Leupold and S.P.~Schneider,
	\emph{{Dispersion relation for hadronic light-by-light scattering: pion
			pole}}, \href{https://doi.org/10.1007/JHEP10(2018)141}{\emph{JHEP} {\bfseries
			10} (2018) 141} [\href{https://arxiv.org/abs/1808.04823}{{\ttfamily
			1808.04823}}].
	
	\bibitem{Gerardin:2019vio}
	A.~G\'erardin, H.B.~Meyer and A.~Nyffeler, \emph{{Lattice calculation of the
			pion transition form factor with $N_f=2+1$ Wilson quarks}},
	\href{https://doi.org/10.1103/PhysRevD.100.034520}{\emph{Phys. Rev. D}
		{\bfseries 100} (2019) 034520}
	[\href{https://arxiv.org/abs/1903.09471}{{\ttfamily 1903.09471}}].
	
	\bibitem{Bijnens:2019ghy}
	J.~Bijnens, N.~Hermansson-Truedsson and A.~Rodr\'\i{}guez-S\'anchez,
	\emph{{Short-distance constraints for the HLbL contribution to the muon
			anomalous magnetic moment}},
	\href{https://doi.org/10.1016/j.physletb.2019.134994}{\emph{Phys. Lett. B}
		{\bfseries 798} (2019) 134994}
	[\href{https://arxiv.org/abs/1908.03331}{{\ttfamily 1908.03331}}].
	
	\bibitem{Colangelo:2019uex}
	G.~Colangelo, F.~Hagelstein, M.~Hoferichter, L.~Laub and P.~Stoffer,
	\emph{{Longitudinal short-distance constraints for the hadronic
			light-by-light contribution to $(g-2)_\mu$ with large-$N_c$ Regge models}},
	\href{https://doi.org/10.1007/JHEP03(2020)101}{\emph{JHEP} {\bfseries 03}
		(2020) 101} [\href{https://arxiv.org/abs/1910.13432}{{\ttfamily
			1910.13432}}].
	
	\bibitem{Blum:2019ugy}
	T.~Blum, N.~Christ, M.~Hayakawa, T.~Izubuchi, L.~Jin, C.~Jung et~al.,
	\emph{{Hadronic Light-by-Light Scattering Contribution to the Muon Anomalous
			Magnetic Moment from Lattice QCD}},
	\href{https://doi.org/10.1103/PhysRevLett.124.132002}{\emph{Phys. Rev. Lett.}
		{\bfseries 124} (2020) 132002}
	[\href{https://arxiv.org/abs/1911.08123}{{\ttfamily 1911.08123}}].
	
	\bibitem{Colangelo:2014qya}
	G.~Colangelo, M.~Hoferichter, A.~Nyffeler, M.~Passera and P.~Stoffer,
	\emph{{Remarks on higher-order hadronic corrections to the muon
			g\ensuremath{-}2}},
	\href{https://doi.org/10.1016/j.physletb.2014.06.012}{\emph{Phys. Lett. B}
		{\bfseries 735} (2014) 90} [\href{https://arxiv.org/abs/1403.7512}{{\ttfamily
			1403.7512}}].
	
	\bibitem{Cherchiglia:2023utd}
	A.L.~Cherchiglia, G.~De~Conto and C.C.~Nishi, \emph{{Connecting (g
			\ensuremath{-} 2)$_{\mu}$ to neutrino mass in the extended neutrinophilic
			2HDM}}, \href{https://doi.org/10.1007/JHEP08(2023)170}{\emph{JHEP} {\bfseries
			08} (2023) 170} [\href{https://arxiv.org/abs/2304.00038}{{\ttfamily
			2304.00038}}].
	
	\bibitem{Crivellin:2021rbq}
	A.~Crivellin and M.~Hoferichter, \emph{{Consequences of chirally enhanced
			explanations of (g \ensuremath{-} 2)$_{\mu}$ for h \textrightarrow{}
			\ensuremath{\mu}\ensuremath{\mu} and Z \textrightarrow{}
			\ensuremath{\mu}\ensuremath{\mu}}},
	\href{https://doi.org/10.1007/JHEP07(2021)135}{\emph{JHEP} {\bfseries 07}
		(2021) 135} [\href{https://arxiv.org/abs/2104.03202}{{\ttfamily
			2104.03202}}].
	
	\bibitem{Crivellin:2018qmi}
	A.~Crivellin, M.~Hoferichter and P.~Schmidt-Wellenburg, \emph{{Combined
			explanations of $(g-2)_{\mu,e}$ and implications for a large muon EDM}},
	\href{https://doi.org/10.1103/PhysRevD.98.113002}{\emph{Phys. Rev. D}
		{\bfseries 98} (2018) 113002}
	[\href{https://arxiv.org/abs/1807.11484}{{\ttfamily 1807.11484}}].
	
	\bibitem{Muong-2:2023cdq}
	{\scshape Muon g-2} collaboration, \emph{{Measurement of the Positive Muon
			Anomalous Magnetic Moment to 0.20 ppm}},
	\href{https://arxiv.org/abs/2308.06230}{{\ttfamily 2308.06230}}.
	
	\bibitem{Aoyama:2020ynm}
	T.~Aoyama et~al., \emph{{The anomalous magnetic moment of the muon in the
			Standard Model}},
	\href{https://doi.org/10.1016/j.physrep.2020.07.006}{\emph{Phys. Rept.}
		{\bfseries 887} (2020) 1} [\href{https://arxiv.org/abs/2006.04822}{{\ttfamily
			2006.04822}}].
	
	\bibitem{CMD-3:2023alj}
	{\scshape CMD-3} collaboration, \emph{{Measurement of the $e^+e^-\to\pi^+\pi^-$
			cross section from threshold to 1.2 GeV with the CMD-3 detector}},
	\href{https://arxiv.org/abs/2302.08834}{{\ttfamily 2302.08834}}.
	
	\bibitem{Jegerlehner:2009ry}
	F.~Jegerlehner and A.~Nyffeler, \emph{{The Muon g-2}},
	\href{https://doi.org/10.1016/j.physrep.2009.04.003}{\emph{Phys. Rept.}
		{\bfseries 477} (2009) 1} [\href{https://arxiv.org/abs/0902.3360}{{\ttfamily
			0902.3360}}].
	
	\bibitem{Lindner:2016bgg}
	M.~Lindner, M.~Platscher and F.S.~Queiroz, \emph{{A Call for New Physics : The
			Muon Anomalous Magnetic Moment and Lepton Flavor Violation}},
	\href{https://doi.org/10.1016/j.physrep.2017.12.001}{\emph{Phys. Rept.}
		{\bfseries 731} (2018) 1} [\href{https://arxiv.org/abs/1610.06587}{{\ttfamily
			1610.06587}}].
	
	\bibitem{Athron:2021iuf}
	P.~Athron, C.~Bal\'azs, D.H.J.~Jacob, W.~Kotlarski, D.~St\"ockinger and
	H.~St\"ockinger-Kim, \emph{{New physics explanations of a$_{\mu}$ in light of
			the FNAL muon g \ensuremath{-} 2 measurement}},
	\href{https://doi.org/10.1007/JHEP09(2021)080}{\emph{JHEP} {\bfseries 09}
		(2021) 080} [\href{https://arxiv.org/abs/2104.03691}{{\ttfamily
			2104.03691}}].
	
	\bibitem{Bhattacharya:2015qpa}
	S.~Bhattacharya, N.~Sahoo and N.~Sahu, \emph{{Minimal vectorlike leptonic dark
			matter and signatures at the LHC}},
	\href{https://doi.org/10.1103/PhysRevD.93.115040}{\emph{Phys. Rev. D}
		{\bfseries 93} (2016) 115040}
	[\href{https://arxiv.org/abs/1510.02760}{{\ttfamily 1510.02760}}].
	
	\bibitem{Bhattacharya:2018fus}
	S.~Bhattacharya, P.~Ghosh, N.~Sahoo and N.~Sahu, \emph{{Mini Review on
			Vector-Like Leptonic Dark Matter, Neutrino Mass, and Collider Signatures}},
	\href{https://doi.org/10.3389/fphy.2019.00080}{\emph{Front. in Phys.}
		{\bfseries 7} (2019) 80} [\href{https://arxiv.org/abs/1812.06505}{{\ttfamily
			1812.06505}}].
	
	\bibitem{Banerjee:2016hsk}
	S.~Banerjee, S.~Matsumoto, K.~Mukaida and Y.-L.S.~Tsai, \emph{{WIMP Dark Matter
			in a Well-Tempered Regime: A case study on Singlet-Doublets Fermionic WIMP}},
	\href{https://doi.org/10.1007/JHEP11(2016)070}{\emph{JHEP} {\bfseries 11}
		(2016) 070} [\href{https://arxiv.org/abs/1603.07387}{{\ttfamily
			1603.07387}}].
	
	\bibitem{DuttaBanik:2018emv}
	A.~Dutta~Banik, A.K.~Saha and A.~Sil, \emph{{Scalar assisted singlet doublet
			fermion dark matter model and electroweak vacuum stability}},
	\href{https://doi.org/10.1103/PhysRevD.98.075013}{\emph{Phys. Rev. D}
		{\bfseries 98} (2018) 075013}
	[\href{https://arxiv.org/abs/1806.08080}{{\ttfamily 1806.08080}}].
	
	\bibitem{Horiuchi:2016tqw}
	S.~Horiuchi, O.~Macias, D.~Restrepo, A.~Rivera, O.~Zapata and H.~Silverwood,
	\emph{{The Fermi-LAT gamma-ray excess at the Galactic Center in the
			singlet-doublet fermion dark matter model}},
	\href{https://doi.org/10.1088/1475-7516/2016/03/048}{\emph{JCAP} {\bfseries
			03} (2016) 048} [\href{https://arxiv.org/abs/1602.04788}{{\ttfamily
			1602.04788}}].
	
	\bibitem{Restrepo:2015ura}
	D.~Restrepo, A.~Rivera, M.~S\'anchez-Pel\'aez, O.~Zapata and W.~Tangarife,
	\emph{{Radiative Neutrino Masses in the Singlet-Doublet Fermion Dark Matter
			Model with Scalar Singlets}},
	\href{https://doi.org/10.1103/PhysRevD.92.013005}{\emph{Phys. Rev. D}
		{\bfseries 92} (2015) 013005}
	[\href{https://arxiv.org/abs/1504.07892}{{\ttfamily 1504.07892}}].
	
	\bibitem{Badziak:2017the}
	M.~Badziak, M.~Olechowski and P.~Szczerbiak, \emph{{Is well-tempered neutralino
			in MSSM still alive after 2016 LUX results?}},
	\href{https://doi.org/10.1016/j.physletb.2017.04.059}{\emph{Phys. Lett. B}
		{\bfseries 770} (2017) 226}
	[\href{https://arxiv.org/abs/1701.05869}{{\ttfamily 1701.05869}}].
	
	\bibitem{Betancur:2020fdl}
	A.~Betancur, G.~Palacio and A.~Rivera, \emph{{Inert doublet as multicomponent
			dark matter}},
	\href{https://doi.org/10.1016/j.nuclphysb.2020.115276}{\emph{Nucl. Phys. B}
		{\bfseries 962} (2021) 115276}
	[\href{https://arxiv.org/abs/2002.02036}{{\ttfamily 2002.02036}}].
	
	\bibitem{Abe:2017glm}
	T.~Abe, \emph{{Effect of CP violation in the singlet-doublet dark matter
			model}}, \href{https://doi.org/10.1016/j.physletb.2017.05.048}{\emph{Phys.
			Lett. B} {\bfseries 771} (2017) 125}
	[\href{https://arxiv.org/abs/1702.07236}{{\ttfamily 1702.07236}}].
	
	\bibitem{Abe:2019wku}
	T.~Abe and R.~Sato, \emph{{Current status and future prospects of the
			singlet-doublet dark matter model with CP-violation}},
	\href{https://doi.org/10.1103/PhysRevD.99.035012}{\emph{Phys. Rev. D}
		{\bfseries 99} (2019) 035012}
	[\href{https://arxiv.org/abs/1901.02278}{{\ttfamily 1901.02278}}].
	
	\bibitem{Barman:2019tuo}
	B.~Barman, S.~Bhattacharya, P.~Ghosh, S.~Kadam and N.~Sahu, \emph{{Fermion Dark
			Matter with Scalar Triplet at Direct and Collider Searches}},
	\href{https://doi.org/10.1103/PhysRevD.100.015027}{\emph{Phys. Rev. D}
		{\bfseries 100} (2019) 015027}
	[\href{https://arxiv.org/abs/1902.01217}{{\ttfamily 1902.01217}}].
	
	\bibitem{Calibbi:2018fqf}
	L.~Calibbi, L.~Lopez-Honorez, S.~Lowette and A.~Mariotti,
	\emph{{Singlet-Doublet Dark Matter Freeze-in: LHC displaced signatures versus
			cosmology}}, \href{https://doi.org/10.1007/JHEP09(2018)037}{\emph{JHEP}
		{\bfseries 09} (2018) 037}
	[\href{https://arxiv.org/abs/1805.04423}{{\ttfamily 1805.04423}}].
	
	\bibitem{Fraser:2020dpy}
	K.~Fraser, A.~Parikh and W.L.~Xu, \emph{{A Closer Look at CP-Violating Higgs
			Portal Dark Matter as a Candidate for the GCE}},
	\href{https://doi.org/10.1007/JHEP03(2021)123}{\emph{JHEP} {\bfseries 03}
		(2021) 123} [\href{https://arxiv.org/abs/2010.15129}{{\ttfamily
			2010.15129}}].
	
	\bibitem{Freitas:2015hsa}
	A.~Freitas, S.~Westhoff and J.~Zupan, \emph{{Integrating in the Higgs Portal to
			Fermion Dark Matter}},
	\href{https://doi.org/10.1007/JHEP09(2015)015}{\emph{JHEP} {\bfseries 09}
		(2015) 015} [\href{https://arxiv.org/abs/1506.04149}{{\ttfamily
			1506.04149}}].
	
	\bibitem{Cynolter:2015sua}
	G.~Cynolter, J.~Kov\'acs and E.~Lendvai, \emph{{Doublet\textendash{}singlet
			model and unitarity}},
	\href{https://doi.org/10.1142/S0217732316500139}{\emph{Mod. Phys. Lett. A}
		{\bfseries 31} (2016) 1650013}
	[\href{https://arxiv.org/abs/1509.05323}{{\ttfamily 1509.05323}}].
	
	\bibitem{Calibbi:2015nha}
	L.~Calibbi, A.~Mariotti and P.~Tziveloglou, \emph{{Singlet-Doublet Model: Dark
			matter searches and LHC constraints}},
	\href{https://doi.org/10.1007/JHEP10(2015)116}{\emph{JHEP} {\bfseries 10}
		(2015) 116} [\href{https://arxiv.org/abs/1505.03867}{{\ttfamily
			1505.03867}}].
	
	\bibitem{Abe:2014gua}
	T.~Abe, R.~Kitano and R.~Sato, \emph{{Discrimination of dark matter models in
			future experiments}},
	\href{https://doi.org/10.1103/PhysRevD.91.095004}{\emph{Phys. Rev. D}
		{\bfseries 91} (2015) 095004}
	[\href{https://arxiv.org/abs/1411.1335}{{\ttfamily 1411.1335}}].
	
	\bibitem{Cheung:2013dua}
	C.~Cheung and D.~Sanford, \emph{{Simplified Models of Mixed Dark Matter}},
	\href{https://doi.org/10.1088/1475-7516/2014/02/011}{\emph{JCAP} {\bfseries
			02} (2014) 011} [\href{https://arxiv.org/abs/1311.5896}{{\ttfamily
			1311.5896}}].
	
	\bibitem{Cohen:2011ec}
	T.~Cohen, J.~Kearney, A.~Pierce and D.~Tucker-Smith, \emph{{Singlet-Doublet
			Dark Matter}}, \href{https://doi.org/10.1103/PhysRevD.85.075003}{\emph{Phys.
			Rev. D} {\bfseries 85} (2012) 075003}
	[\href{https://arxiv.org/abs/1109.2604}{{\ttfamily 1109.2604}}].
	
	\bibitem{Dutta:2021uxd}
	M.~Dutta, S.~Bhattacharya, P.~Ghosh and N.~Sahu, \emph{{Majorana Dark Matter
			and~Neutrino Mass in~a~Singlet-Doublet Extension of~the~Standard Model}},
	\href{https://doi.org/10.1007/978-981-19-2354-8_124}{\emph{Springer Proc.
			Phys.} {\bfseries 277} (2022) 685}
	[\href{https://arxiv.org/abs/2106.13857}{{\ttfamily 2106.13857}}].
	
	\bibitem{Ghosh:2023dhj}
	P.~Ghosh and S.~Jeesun, \emph{{Reviving sub-TeV $SU(2)_L$ lepton doublet dark
			matter}}, \href{https://doi.org/10.1140/epjc/s10052-023-12039-z}{\emph{Eur.
			Phys. J. C} {\bfseries 83} (2023) 880}
	[\href{https://arxiv.org/abs/2306.12906}{{\ttfamily 2306.12906}}].
	
	\bibitem{Borah:2021khc}
	D.~Borah, M.~Dutta, S.~Mahapatra and N.~Sahu, \emph{{Lepton anomalous magnetic
			moment with singlet-doublet fermion dark matter in a scotogenic
			U(1)L\ensuremath{\mu}-L\ensuremath{\tau} model}},
	\href{https://doi.org/10.1103/PhysRevD.105.015029}{\emph{Phys. Rev. D}
		{\bfseries 105} (2022) 015029}
	[\href{https://arxiv.org/abs/2109.02699}{{\ttfamily 2109.02699}}].
	
	\bibitem{Bhattacharya:2021ltd}
	S.~Bhattacharya, S.~Jahedi and J.~Wudka, \emph{{Probing heavy charged fermions
			at e$^{+}$e$^{-}$ collider using the optimal observable technique}},
	\href{https://doi.org/10.1007/JHEP05(2022)009}{\emph{JHEP} {\bfseries 05}
		(2022) 009} [\href{https://arxiv.org/abs/2106.02846}{{\ttfamily
			2106.02846}}].
	
	\bibitem{Bhattacharya:2018cgx}
	S.~Bhattacharya, P.~Ghosh and N.~Sahu, \emph{{Multipartite Dark Matter with
			Scalars, Fermions and signatures at LHC}},
	\href{https://doi.org/10.1007/JHEP02(2019)059}{\emph{JHEP} {\bfseries 02}
		(2019) 059} [\href{https://arxiv.org/abs/1809.07474}{{\ttfamily
			1809.07474}}].
	
	\bibitem{Bhattacharya:2016rqj}
	S.~Bhattacharya, B.~Karmakar, N.~Sahu and A.~Sil, \emph{{Flavor origin of dark
			matter and its relation with leptonic nonzero $\theta_{13}$ and Dirac CP
			phase $\delta$}}, \href{https://doi.org/10.1007/JHEP05(2017)068}{\emph{JHEP}
		{\bfseries 05} (2017) 068}
	[\href{https://arxiv.org/abs/1611.07419}{{\ttfamily 1611.07419}}].
	
	\bibitem{Bhattacharya:2017sml}
	S.~Bhattacharya, N.~Sahoo and N.~Sahu, \emph{{Singlet-Doublet Fermionic Dark
			Matter, Neutrino Mass and Collider Signatures}},
	\href{https://doi.org/10.1103/PhysRevD.96.035010}{\emph{Phys. Rev. D}
		{\bfseries 96} (2017) 035010}
	[\href{https://arxiv.org/abs/1704.03417}{{\ttfamily 1704.03417}}].
	
	\bibitem{Dutta:2020xwn}
	M.~Dutta, S.~Bhattacharya, P.~Ghosh and N.~Sahu, \emph{{Singlet-Doublet
			Majorana Dark Matter and Neutrino Mass in a minimal Type-I Seesaw Scenario}},
	\href{https://doi.org/10.1088/1475-7516/2021/03/008}{\emph{JCAP} {\bfseries
			03} (2021) 008} [\href{https://arxiv.org/abs/2009.00885}{{\ttfamily
			2009.00885}}].
	
	\bibitem{Konar:2020wvl}
	P.~Konar, A.~Mukherjee, A.K.~Saha and S.~Show, \emph{{Linking pseudo-Dirac dark
			matter to radiative neutrino masses in a singlet-doublet scenario}},
	\href{https://doi.org/10.1103/PhysRevD.102.015024}{\emph{Phys. Rev. D}
		{\bfseries 102} (2020) 015024}
	[\href{https://arxiv.org/abs/2001.11325}{{\ttfamily 2001.11325}}].
	
	\bibitem{Konar:2020vuu}
	P.~Konar, A.~Mukherjee, A.K.~Saha and S.~Show, \emph{{A dark clue to seesaw and
			leptogenesis in a pseudo-Dirac singlet doublet scenario with (non)standard
			cosmology}}, \href{https://doi.org/10.1007/JHEP03(2021)044}{\emph{JHEP}
		{\bfseries 03} (2021) 044}
	[\href{https://arxiv.org/abs/2007.15608}{{\ttfamily 2007.15608}}].
	
	\bibitem{Borah:2022zim}
	D.~Borah, S.~Mahapatra and N.~Sahu, \emph{{Singlet-doublet fermion origin of
			dark matter, neutrino mass and W-mass anomaly}},
	\href{https://doi.org/10.1016/j.physletb.2022.137196}{\emph{Phys. Lett. B}
		{\bfseries 831} (2022) 137196}
	[\href{https://arxiv.org/abs/2204.09671}{{\ttfamily 2204.09671}}].
	
	\bibitem{Borah:2021rbx}
	D.~Borah, M.~Dutta, S.~Mahapatra and N.~Sahu, \emph{{Singlet-doublet
			self-interacting dark matter and radiative neutrino mass}},
	\href{https://doi.org/10.1103/PhysRevD.105.075019}{\emph{Phys. Rev. D}
		{\bfseries 105} (2022) 075019}
	[\href{https://arxiv.org/abs/2112.06847}{{\ttfamily 2112.06847}}].
	
	\bibitem{Abazajian:2019eic}
	K.~Abazajian et~al., \emph{{CMB-S4 Science Case, Reference Design, and Project
			Plan}},  \href{https://arxiv.org/abs/1907.04473}{{\ttfamily 1907.04473}}.
	
	\bibitem{Barbieri:2004qk}
	R.~Barbieri, A.~Pomarol, R.~Rattazzi and A.~Strumia, \emph{{Electroweak
			symmetry breaking after LEP-1 and LEP-2}},
	\href{https://doi.org/10.1016/j.nuclphysb.2004.10.014}{\emph{Nucl. Phys. B}
		{\bfseries 703} (2004) 127}
	[\href{https://arxiv.org/abs/hep-ph/0405040}{{\ttfamily hep-ph/0405040}}].
	
	\bibitem{MEG:2016leq}
	{\scshape MEG} collaboration, \emph{{Search for the lepton flavour violating
			decay $\mu ^+ \rightarrow \mathrm {e}^+ \gamma $ with the full dataset of the
			MEG experiment}},
	\href{https://doi.org/10.1140/epjc/s10052-016-4271-x}{\emph{Eur. Phys. J. C}
		{\bfseries 76} (2016) 434}
	[\href{https://arxiv.org/abs/1605.05081}{{\ttfamily 1605.05081}}].
	
	\bibitem{Belanger:2014vza}
	G.~B\'elanger, F.~Boudjema, A.~Pukhov and A.~Semenov, \emph{{micrOMEGAs4.1: two
			dark matter candidates}},
	\href{https://doi.org/10.1016/j.cpc.2015.03.003}{\emph{Comput. Phys. Commun.}
		{\bfseries 192} (2015) 322}
	[\href{https://arxiv.org/abs/1407.6129}{{\ttfamily 1407.6129}}].
	
	\bibitem{Semenov:2008jy}
	A.~Semenov, \emph{{LanHEP: A Package for the automatic generation of Feynman
			rules in field theory. Version 3.0}},
	\href{https://doi.org/10.1016/j.cpc.2008.10.012}{\emph{Comput. Phys. Commun.}
		{\bfseries 180} (2009) 431}
	[\href{https://arxiv.org/abs/0805.0555}{{\ttfamily 0805.0555}}].
	
	\bibitem{Goodman:1984dc}
	M.W.~Goodman and E.~Witten, \emph{{Detectability of Certain Dark Matter
			Candidates}}, \href{https://doi.org/10.1103/PhysRevD.31.3059}{\emph{Phys.
			Rev. D} {\bfseries 31} (1985) 3059}.
	
	\bibitem{Essig:2007az}
	R.~Essig, \emph{{Direct Detection of Non-Chiral Dark Matter}},
	\href{https://doi.org/10.1103/PhysRevD.78.015004}{\emph{Phys. Rev. D}
		{\bfseries 78} (2008) 015004}
	[\href{https://arxiv.org/abs/0710.1668}{{\ttfamily 0710.1668}}].
	
	\bibitem{Bertone:2004pz}
	G.~Bertone, D.~Hooper and J.~Silk, \emph{{Particle dark matter: Evidence,
			candidates and constraints}},
	\href{https://doi.org/10.1016/j.physrep.2004.08.031}{\emph{Phys. Rept.}
		{\bfseries 405} (2005) 279}
	[\href{https://arxiv.org/abs/hep-ph/0404175}{{\ttfamily hep-ph/0404175}}].
	
	\bibitem{Alarcon:2012nr}
	J.M.~Alarcon, L.S.~Geng, J.~Martin~Camalich and J.A.~Oller, \emph{{The
			strangeness content of the nucleon from effective field theory and
			phenomenology}},
	\href{https://doi.org/10.1016/j.physletb.2014.01.065}{\emph{Phys. Lett. B}
		{\bfseries 730} (2014) 342}
	[\href{https://arxiv.org/abs/1209.2870}{{\ttfamily 1209.2870}}].
	
	\bibitem{Hoferichter:2017olk}
	M.~Hoferichter, P.~Klos, J.~Men\'endez and A.~Schwenk, \emph{{Improved limits
			for Higgs-portal dark matter from LHC searches}},
	\href{https://doi.org/10.1103/PhysRevLett.119.181803}{\emph{Phys. Rev. Lett.}
		{\bfseries 119} (2017) 181803}
	[\href{https://arxiv.org/abs/1708.02245}{{\ttfamily 1708.02245}}].
	
	\bibitem{Ibarra:2016dlb}
	A.~Ibarra, C.E.~Yaguna and O.~Zapata, \emph{{Direct Detection of Fermion Dark
			Matter in the Radiative Seesaw Model}},
	\href{https://doi.org/10.1103/PhysRevD.93.035012}{\emph{Phys. Rev. D}
		{\bfseries 93} (2016) 035012}
	[\href{https://arxiv.org/abs/1601.01163}{{\ttfamily 1601.01163}}].
	
	\bibitem{XENON:2023cxc}
	{\scshape XENON} collaboration, \emph{{First Dark Matter Search with Nuclear
			Recoils from the XENONnT Experiment}},
	\href{https://doi.org/10.1103/PhysRevLett.131.041003}{\emph{Phys. Rev. Lett.}
		{\bfseries 131} (2023) 041003}
	[\href{https://arxiv.org/abs/2303.14729}{{\ttfamily 2303.14729}}].
	
	\bibitem{LZ:2022lsv}
	{\scshape LZ} collaboration, \emph{{First Dark Matter Search Results from the
			LUX-ZEPLIN (LZ) Experiment}},
	\href{https://doi.org/10.1103/PhysRevLett.131.041002}{\emph{Phys. Rev. Lett.}
		{\bfseries 131} (2023) 041002}
	[\href{https://arxiv.org/abs/2207.03764}{{\ttfamily 2207.03764}}].
	
	\bibitem{DARWIN:2016hyl}
	{\scshape DARWIN} collaboration, \emph{{DARWIN: towards the ultimate dark
			matter detector}},
	\href{https://doi.org/10.1088/1475-7516/2016/11/017}{\emph{JCAP} {\bfseries
			11} (2016) 017} [\href{https://arxiv.org/abs/1606.07001}{{\ttfamily
			1606.07001}}].
	
	\bibitem{Mangano:2005cc}
	G.~Mangano, G.~Miele, S.~Pastor, T.~Pinto, O.~Pisanti and P.D.~Serpico,
	\emph{{Relic neutrino decoupling including flavor oscillations}},
	\href{https://doi.org/10.1016/j.nuclphysb.2005.09.041}{\emph{Nucl. Phys. B}
		{\bfseries 729} (2005) 221}
	[\href{https://arxiv.org/abs/hep-ph/0506164}{{\ttfamily hep-ph/0506164}}].
	
	\bibitem{Grohs:2015tfy}
	E.~Grohs, G.M.~Fuller, C.T.~Kishimoto, M.W.~Paris and A.~Vlasenko,
	\emph{{Neutrino energy transport in weak decoupling and big bang
			nucleosynthesis}},
	\href{https://doi.org/10.1103/PhysRevD.93.083522}{\emph{Phys. Rev. D}
		{\bfseries 93} (2016) 083522}
	[\href{https://arxiv.org/abs/1512.02205}{{\ttfamily 1512.02205}}].
	
	\bibitem{deSalas:2016ztq}
	P.F.~de~Salas and S.~Pastor, \emph{{Relic neutrino decoupling with flavour
			oscillations revisited}},
	\href{https://doi.org/10.1088/1475-7516/2016/07/051}{\emph{JCAP} {\bfseries
			07} (2016) 051} [\href{https://arxiv.org/abs/1606.06986}{{\ttfamily
			1606.06986}}].
	
	\bibitem{Cielo:2023bqp}
	M.~Cielo, M.~Escudero, G.~Mangano and O.~Pisanti, \emph{{Neff in the Standard
			Model at NLO is 3.043}},  \href{https://arxiv.org/abs/2306.05460}{{\ttfamily
			2306.05460}}.
	
	\bibitem{Akita:2020szl}
	K.~Akita and M.~Yamaguchi, \emph{{A precision calculation of relic neutrino
			decoupling}},
	\href{https://doi.org/10.1088/1475-7516/2020/08/012}{\emph{JCAP} {\bfseries
			08} (2020) 012} [\href{https://arxiv.org/abs/2005.07047}{{\ttfamily
			2005.07047}}].
	
	\bibitem{Froustey:2020mcq}
	J.~Froustey, C.~Pitrou and M.C.~Volpe, \emph{{Neutrino decoupling including
			flavour oscillations and primordial nucleosynthesis}},
	\href{https://doi.org/10.1088/1475-7516/2020/12/015}{\emph{JCAP} {\bfseries
			12} (2020) 015} [\href{https://arxiv.org/abs/2008.01074}{{\ttfamily
			2008.01074}}].
	
	\bibitem{Bennett:2020zkv}
	J.J.~Bennett, G.~Buldgen, P.F.~De~Salas, M.~Drewes, S.~Gariazzo, S.~Pastor
	et~al., \emph{{Towards a precision calculation of $N_{\rm eff}$ in the
			Standard Model II: Neutrino decoupling in the presence of flavour
			oscillations and finite-temperature QED}},
	\href{https://doi.org/10.1088/1475-7516/2021/04/073}{\emph{JCAP} {\bfseries
			04} (2021) 073} [\href{https://arxiv.org/abs/2012.02726}{{\ttfamily
			2012.02726}}].
	
	\bibitem{SPT-3G:2019sok}
	{\scshape SPT-3G} collaboration, \emph{{Particle Physics with the Cosmic
			Microwave Background with SPT-3G}},
	\href{https://doi.org/10.1088/1742-6596/1468/1/012008}{\emph{J. Phys. Conf.
			Ser.} {\bfseries 1468} (2020) 012008}
	[\href{https://arxiv.org/abs/1911.08047}{{\ttfamily 1911.08047}}].
	
	\bibitem{Luo:2020sho}
	X.~Luo, W.~Rodejohann and X.-J.~Xu, \emph{{Dirac neutrinos and $N_{{\rm
					eff}}$}}, \href{https://doi.org/10.1088/1475-7516/2020/06/058}{\emph{JCAP}
		{\bfseries 06} (2020) 058}
	[\href{https://arxiv.org/abs/2005.01629}{{\ttfamily 2005.01629}}].
	
	\bibitem{Luo:2020fdt}
	X.~Luo, W.~Rodejohann and X.-J.~Xu, \emph{{Dirac neutrinos and N$_{eff}$. Part
			II. The freeze-in case}},
	\href{https://doi.org/10.1088/1475-7516/2021/03/082}{\emph{JCAP} {\bfseries
			03} (2021) 082} [\href{https://arxiv.org/abs/2011.13059}{{\ttfamily
			2011.13059}}].
	
\end{thebibliography}

\providecommand{\href}[2]{#2}\begingroup\raggedright\endgroup

\end{document}